\documentclass[sigplan,10pt,screen]{acmart}
\renewcommand\footnotetextcopyrightpermission[1]{}
\usepackage{pifont}
\usepackage{amsmath}
\usepackage{xspace}
\usepackage{multirow}
\usepackage{algorithm}
\usepackage{graphicx}
\usepackage{mathtools}
\usepackage{amsthm}
\usepackage{wrapfig}
\usepackage{fancyhdr}
\usepackage{bm}
\usepackage{rotating}
\usepackage{listings}
\usepackage{xcolor}
\usepackage{enumitem}
\usepackage[noend]{algorithmic}

\usepackage{microtype}

\definecolor{dkgreen}{rgb}{0,0.6,0}
\definecolor{carrotorange}{rgb}{0.93, 0.57, 0.13}

\edef\oldtt{\ttdefault}
\usepackage[scaled]{beramono}
\usepackage[T1]{fontenc}
\renewcommand{\ttdefault}{\oldtt}

\newcommand{\todo}[1]{\textcolor{black}{#1}}
\newcommand{\red}[1]{\textcolor{black}{#1}}
\newcommand{\blue}[1]{\textcolor{black}{#1}}

\newcommand{\redasplos}[1]{\textcolor{black}{#1}}

\newcommand{\revision}[1]{\textcolor{black}{#1}}

\newcommand{\maxspeedupautomine}{$827\times$\xspace}
\newcommand{\maxspeedupperegrine}{$575\times$\xspace}
\newcommand{\maxspeedupfractal}{$42,143\times$\xspace}
\newcommand{\maxspeeduppangolin}{$328\times$\xspace}
\newcommand{\maxspeeduprstream}{$882,667\times$\xspace}
\newcommand{\maxspeeduparabesque}{$72,143\times$\xspace}
\newcommand{\maxspeedupgraphpi}{$62.8\times$\xspace}

\begin{document}

\title{DwarvesGraph: A High-Performance Graph Mining System with Pattern Decomposition}

\author{Jingji Chen}
\email{chen3385@purdue.edu}
\affiliation{%
  \institution{Purdue University}
  \city{West Lafayette, IN}
  \country{USA}}

\author{Xuehai Qian}
\email{qian214@purdue.edu}
\affiliation{%
  \institution{Purdue University}
  \city{West Lafayette, IN}
  \country{USA}}

\newcommand{\proj}{DwarvesGraph\xspace}

\begin{abstract}

Graph pattern mining (GPM) is an important application that 
identifies structures from graphs. 
Despite the recent progress,
the performance gap between the state-of-the-art GPM 
systems and an efficient algorithm---pattern decomposition---is still at least an order of magnitude. 
This paper clears the fundamental obstacles of adopting 
pattern decomposition to a GPM system.

First, the performance of pattern decomposition algorithms
depends on how to decompose the whole pattern into 
subpatterns. 
The original method
performs complexity analysis of algorithms for 
different choices,
and selects the one with the lowest
complexity upper bound. 
Clearly, this approach is not feasible for average or 
even expert users. 
To solve this problem, we develop a GPM compiler with 
conventional and GPM-specific optimizations to
generate algorithms for different decomposition choices, which
are evaluated based on an accurate cost model.
The executable of the GPM task is obtained from the 
algorithm with the best performance. 
Second, we propose a novel partial-embedding API that 
is sufficient to construct advanced GPM applications 
while preserving pattern
decomposition algorithm advantages. 
Compared to state-of-the-art systems, 
our new GPM system, \proj, developed based on the ideas, 
reduces the execution time of GPM on large graphs and patterns 
from days to a few hours with low programming effort. 


\end{abstract}

\settopmatter{printfolios=true}
\settopmatter{printacmref=false}
\maketitle
\pagestyle{plain}

\section{Introduction}
\label{sec:introduction}

Graph pattern mining (GPM)~\cite{teixeira2015arabesque} is an
important workload~\cite{ribeiro2019survey,shaw1999methods,uddin2013dyad,duma2014network}
with the goal of finding all subgraphs, known as {\em embeddings}, in the input graph that ``match'', i.e., {\em isomorphic} to user-specified patterns.
GPM has important applications in functional modules discovery~\cite{schmidt2011efficient}, biochemical structures mining~\cite{ma2009insights} and anomaly detection~\cite{becchetti2008efficient} and 
many others.
The key challenge of GPM is to enumerate a large 
number of subgraphs.
In WikiVote, a small graph with 7k vertices, the number of embeddings matching the 5-chain pattern
can reach 71 billion based on our experiment. 
For frequent subgraph mining (FSM), 
mining all frequent size-3 patterns with certain criteria 
on a median-size LiveJournal graph by Perergrine~\cite{jamshidi2020peregrine}, one of the recent systems, can take more than 12 hours. 

Due to the importance of GPM for various applications, 
several domain-specific GPM systems
have been proposed in recent years~\cite{teixeira2015arabesque,wang2018rstream,mawhirter2019automine,jamshidi2020peregrine,chen2020pangolin,shi2020graphpi,dias2019fractal,chen2021sandslash}. 
The intuitive APIs, i.e., 
requesting the count of a given pattern
or defining operations for each identified embedding of 
a pattern returned from the system
in user-defined functions (UDFs),
allow non-expert users to construct GPM applications
while achieving high performance. 
We observe that 
the performance gap between the state-of-the-art systems and 
the efficient algorithm is still large.
For example,
using one core, the fastest execution time (617.2s) of mining 
\todo{all size-5}
patterns from the WikiVote graph using GraphPi~\cite{shi2020graphpi} is 48.6$\times$ slower than the  manually-optimized pattern decomposition algorithm~\cite{pinar2017escape} (12.7s).


\begin{figure}[htbp]
    \centering
    \includegraphics[width=.9\linewidth]{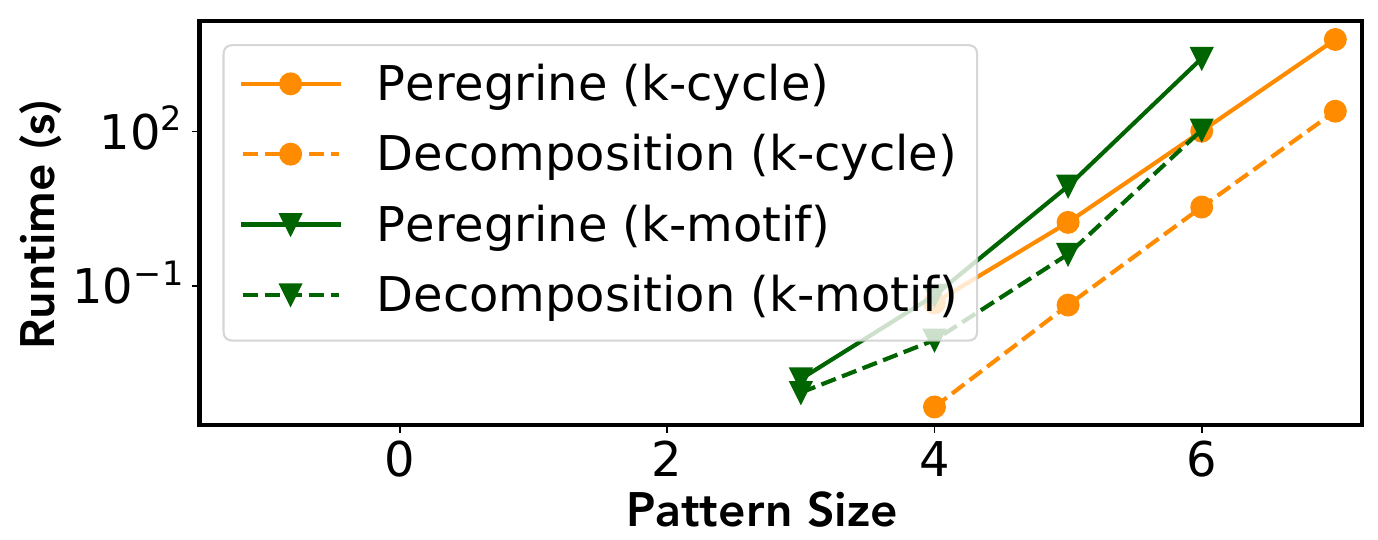}
    \caption{\revision{Pattern Size vs. Runtime}}
\label{fig:cost_growing}
\end{figure}

The key idea of {\em pattern decomposition} is to 
decompose a target pattern into 
smaller subpatterns, and then compute the 
count of each.
The count of the target pattern can be calculated 
using the subpattern counts
with low additional cost. 
The method is very fast 
because, empirically, the embedding enumeration cost increases rapidly with pattern size. 
Figure~\ref{fig:cost_growing} shows
the execution time of Peregrine~\cite{jamshidi2020peregrine} when counting $k$-motif (all size-$k$ patterns) and $k$-cycle (a cycle-structure pattern with $k$ vertices/edges)
on the EmailEuCore graph~\cite{yin2017local,leskovec2007graph}
with increasing pattern size ($k$). 
In comparison, we show the execution time
of an implementation based on the 
pattern decomposition algorithm.
There are two fundamental 
obstacles to adopt the algorithm to an existing GPM
system.

{\bf Obstacle 1: APIs incompatible for pattern decomposition.}
A GPM system should not only perform pattern counting but 
also allow application-specific operations to be expressed
in the user-defined functions (UDFs) for each pattern 
embedding identified by the system. 
Such API is problematic for pattern decomposition:
designed for pattern counting, the
algorithm {counts without materializing} 
any whole pattern 
embedding. However, 
the current API {\em forces the system to materialize and returns 
each whole pattern embedding to UDFs}.
As a result, the algorithmic and performance advantages
are largely wiped out.
Thus, the {\em fundamental problem} is to seek an API that 
can intrinsically work with pattern decomposition and 
preserve most of its advantages, while still being sufficient
to express applications and easy to use. 

{\bf Obstacle 2: Automating pattern decomposition.}
Pattern decomposition algorithm 
decomposes the target pattern by selecting 
a {\em vertex cutting set $V_C$}---removing the 
vertices in the set will break the pattern 
graph $p$ into $K$ connected components. 
The $K$ subpatterns, i.e., $p_1,...,p_K$, are generated by 
merging the vertex cutting set with each of 
the $K$ components.
The algorithm identifies the subgraphs matching $V_C$, denoted
as $e_C$.
For each $e_C$, it counts the {\em number} of subgraphs containing
$e_C$ that can match each subpattern $p_i$, denoted as $M_i$.
With certain exceptions, $M_1 \times ... \times M_K$ is the 
number of embeddings matching $p$ containing $e_C$.
This procedure is illustrated in Figure~\ref{fig:example_pattern_graph}.
We see that different choices of $V_C$ lead to different 
sets of subpatterns and different algorithms. 
The original paper~\cite{pinar2017escape}
performs {\em complexity analysis} of {\em manually-optimized}
algorithms for 
different $V_C$ choices on specific patterns,
and selects the one with the lowest
complexity upper bound. 
This approach is not feasible for average or 
even expert users. 
In a GPM system, the $V_C$ selection process should be {\em automated}. 

\revision{Specifically, the system should search good implementations,
instead of just applying the original pattern decomposition algorithm,
which can perform even worse. A bad selection can be more than one order of magnitude slower than a 
good selection.
For example, for pattern p5 in Figure~\ref{fig:cost_compare} (a), the worst selection is more than $60\times$ slower than the best one on the EmailEuCore graph.
The key challenge is predicting the performance of the implementations 
with an accurate cost model}. 


\begin{figure}[htbp]
    \centering
    \includegraphics[width=\linewidth]{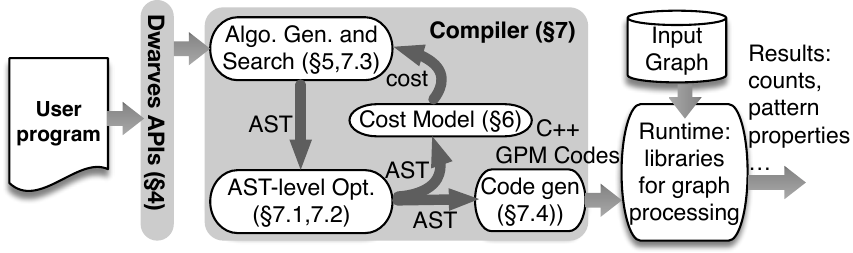} 
    \caption{\proj System Overview}
    \label{fig:overall_architecture}
\end{figure}

{\bf Our solution: \proj---a new compilation-based GPM system 
with a novel API that automates algorithm selection using
the accurate cost model.}
We build the \proj compiler 
that can {\em automatically} generate algorithms for 
{\em arbitrary} patterns with different $V_C$ and pattern 
vertex matching orders, i.e., the order that the vertices 
in the pattern graph are enumerated.
Specifically, \proj compiler generates
the intermediate representation, i.e., abstract syntax tree (AST), of algorithm candidates, 
which are evaluated using the cost model.
For each AST, the compiler performs not only conventional 
optimizations such as Loop Invariant Code Motion (LICM)
and Common Subexpression Elimination (CSE) but also 
a novel {\em pattern-aware loop
rewriting optimization} to eliminate 
redundant computation.
The system overview is shown in Figure~\ref{fig:overall_architecture}.
We make
three novel contributions that
{\em bridge pattern decomposition algorithm and system}.

$\bullet$ {\bf Partial-embedding API: the natural API for 
pattern decomposition}. 
Instead of forcing the system to return whole embedding of 
pattern $p$, the partial-embedding API only requires the system 
to return ``partial'' embeddings for $p_1$,...,$p_K$---{\em any}
valid pattern decomposition. 
It avoids the overhead of materializing whole pattern embedding
while still allowing UDFs to be specified on partial embeddings. 
We find that this API can correctly express 
advanced GPM applications such as Frequent Subgraph Mining (FSM)
as long as the system guarantees two {\em properties}
when passing partial embeddings. 
Partial-embedding API is an {\em elegant} abstraction because
it is inherently compatible with pattern decomposition while 
{\em not requiring users to determine how pattern is 
decomposed}---the responsibility of the \proj compiler.

$\bullet$ {\bf Generalized pattern decomposition algorithm. }
The current pattern decomposition algorithms only support
{\em pattern counting for patterns with no more than five vertices}~\cite{pinar2017escape}, but the \proj compiler
needs to generate algorithms for {\em arbitrary} patterns. 
To this end, we propose a generalized
decomposition algorithm that returns partial embeddings
during execution.

$\bullet$ {\bf Accurate cost models: capturing real-world graph characteristics}. 
The pattern enumeration procedure can be expressed as 
nested loops.
Based on the cost model, we need to 
estimate the number of
iterations that a loop will execute, which depends on the input
graph. For example, for a graph with $n$ vertices,
the loop to process the 1-hop neighbors can be estimated to 
execute $np$ iterations, where $p$ is the probability that 
a vertex is connected to another vertex.
It is the method adopted by AutoMine~\cite{mawhirter2019automine}, i.e., a 
recent compilation-based GPM system with {\em no support for
pattern decomposition} and used as one of the baselines.
We propose two new cost models that significantly 
improve the accuracy.

The {\em locality-aware} cost model is a simple extension
to AutoMine's method that 
assigns a higher connection probability
if two vertices are within $k$-hop, where $k$ can 
be specified as a threshold.
It coarsely considers the real graph properties 
by adjusting the probability.
The second {\em approximate-mining based} cost model
is based on a {\em key observation}:
we can {\em estimate the number of loop 
iterations at a loop level by the approximate
count of the corresponding pattern reaching that level}. 
Refer to Figure~\ref{cost_model_insights},
the 4-vertex pattern can be enumerated with 
matching order
\ding{182} (A,B,C,D) or \ding{183} (D,B,C,A) among others possibilities.
The number of iterations for the loop
to match vertex C can be estimated by the {\em count} of
smaller pattern 
(A,B,C) for \ding{182} and (D,B,C) for 
\ding{183}.
We develop a {\em fast} method to obtain the 
{\em approximate} and {\em relative} count of all patterns
in input graph up to a certain size
\revision{with random sampling and approximate pattern mining~\cite{iyer2018asap}}.

{\bf Alternative solutions. }
A few systems~\cite{chen2020pangolin,chen2021sandslash},
provide low-level but more flexible APIs that
allow users to construct sophisticated algorithms {\em manually}.
The low-level APIs
of Sandslash~\cite{chen2021sandslash} can express pattern decomposition
algorithms. 
However, the manual 
implementation based on the APIs makes it {\em almost the 
same as constructing the native algorithms for 
specific patterns}. 

{\bf Evaluation highlights.} 
The implementation of \proj has about 10,000 lines of codes. We perform comprehensive performance
evaluation comparing \proj with recent
GPM systems---Arabesque~\cite{teixeira2015arabesque},
RStream~\cite{wang2018rstream}, our own implementation of AutoMine~\cite{mawhirter2019automine} (AutoMineInHouse) which
is not open sourced~\footnote{AutoMineInHouse achieves comparable  performance with the original implementation reported in 
~\cite{mawhirter2019automine}, detailed numbers are reported in Table~\ref{tab:compare_automine}.}, Peregrine~\cite{jamshidi2020peregrine}, Pangolin~\cite{chen2020pangolin}, Fractal~\cite{dias2019fractal}, and 
GraphPi~\cite{shi2020graphpi}.
Our results show that \proj 
can be {\em up to \maxspeedupautomine and \maxspeedupperegrine faster than
AutoMineInHouse and Peregrine}, respectively. 
For {\em large graphs}, on Friendster and RMAT-100M, two large graphs with more than one billion edges, \proj reduces the execution time of 4-motif counting {\em from tens of hours to less than two hours compared to Peregrine}.
For {\em large patterns},
AutoMineInHouse fails to mine all 6-motif patterns in {\em one week} on a small graph with 7k vertices; while \proj can finish the same task with roughly {\em one hour} on a median-size graph with 3.8M vertices. 
Most importantly, \proj effectively closes the gap 
between the GPM system and efficient native algorithms.
With multiple threads, \proj achieves better performance.

{\bf Sources of performance improvements. }
The primary reason for the drastic performance improvements is 
due to the better algorithm, but the proposed novel techniques
{\em bring algorithmic advantages to a general GPM system}. 
\proj fully automates the algorithm 
generation, optimization, and selection
from the search space, ensuring the ease-of-use and 
high performance at the same time.
We believe that \proj is a significant
advance over the state-of-the-art of GPM systems.

\section{Graph Pattern Mining Background}
\label{sec:background}

\subsection{Definitions}
\label{prob_def}

\begin{itemize}[leftmargin=*]
    \item {\bf Graph. } $G=(V,E)$ contains a vertex set $V$ and edge set $E$. $E$ is an subset of $V\times V$ and $(u,v)\in E$ if. $u$ and $v$ are connected by an edge. 
    \item {\bf Labeled graph. } Graph vertices may have labels captured by a mapping $f_L:V\rightarrow L$ where $L$ is the set of vertex labels.
    \item {\bf Edge induced subgraph. } It contains a subset of the edges and all corresponding vertices. Formally, $G'=(V',E')$
    is a edge-induced subgraph iff. $E' \in E$ 
    and $V'= \{v \in V | (v,u) \in E' \text{ for some }u\}$.
    \item {\bf Vertex induced subgraph. } It contains a subset of the vertices and all edges induced by them. Formally, $G'=(V',E')$
    is a vertex-induced subgraph iff. $V' \in V$ 
    and $E'= \{(u,v) \in E | u,v \in V'\}$. 
    \item {\bf Isomorphism.} Two graphs $G_0=(V_0, E_0)$ and $G_1=(V_1, E_1)$ are {\em isomorphic} iff. there exists a one-to-one mapping $f: V_0\rightarrow V_1$ such that $(u, v)\in E_0 \iff (f(u), f(v))\in E_1$.
    \item {\bf Graph pattern mining (GPM).} 
    It takes an undirected graph as input, enumerate all its subgraphs that are isomorphic to a given pattern graph, and process them to gather information. When the pattern graph has vertex labels, only the isomorphic subgraphs with matching labels are considered.
    \item {\bf Embedding. } In GPM, it refers to 
    the subgraphs that are isomorphic to the pattern graph.
    \item {\bf Edge (vertex) induced embedding. }
    The embedding is an edge (vertex) induced subgraph.
\end{itemize}    

Figure~\ref{fig:gpm_background} shows an example
of embedding of the pattern graph in input graph. 


\subsection{Pattern Enumeration Method and Optimization}
\label{method_opt}

\begin{figure}[htbp]
    \centering
    \includegraphics[width=.9\linewidth]{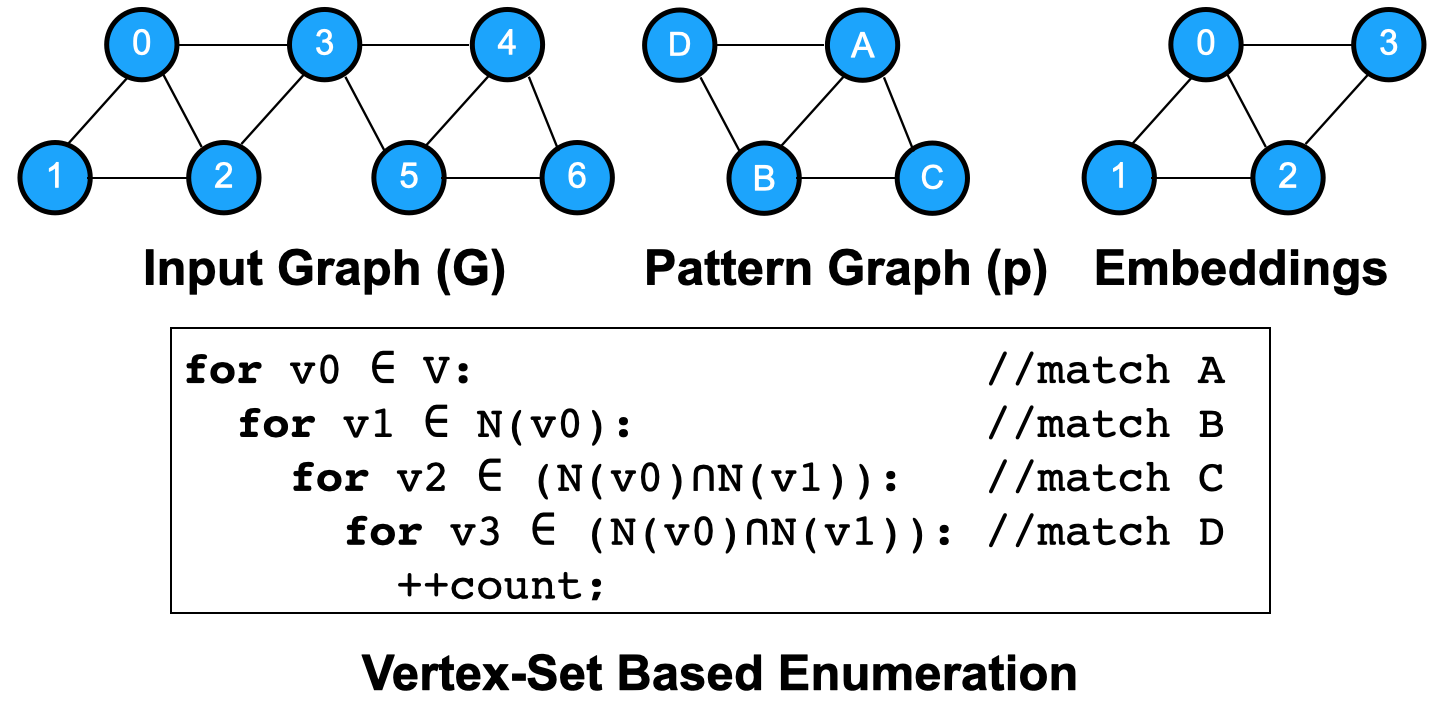}
    \caption{GPM Example and Implementation}
    \label{fig:gpm_background}
\end{figure}

Recent GPM systems~\cite{mawhirter2019automine,jamshidi2020peregrine,shi2020graphpi,mawhirter2021graphzero} use a pattern-aware 
vertex-set-based method to construct the embeddings
based on the pattern graph. 
It significantly outperforms pattern-oblivious method 
used in early systems~\cite{teixeira2015arabesque,wang2018rstream,dias2019fractal} that enumerate all 
subgraphs and perform the expensive isomorphic check. 
Thus, \proj adopts the vertex-set-based method, 
which uses nested loops to extend subgraphs incrementally
from a single vertex until reaching the embedding of 
the pattern graph. 
Figure~\ref{fig:gpm_background} shows one possible 
implementation of
nested loops to enumerate embeddings for the pattern, 
where $N(v)$ refers to the vertex set containing all neighbors of $v$.
It also illustrate the concept of {\em pattern vertex matching order}.
A different matching order, i.e., (D,B,C,A), will
lead to different nested loops that perform the 
equivalent enumeration. 
A {\em cost model} can be utilized to search for
pattern vertex matching orders with high performance.

\begin{figure}[htbp]
    \centering
    \includegraphics[width=\linewidth]{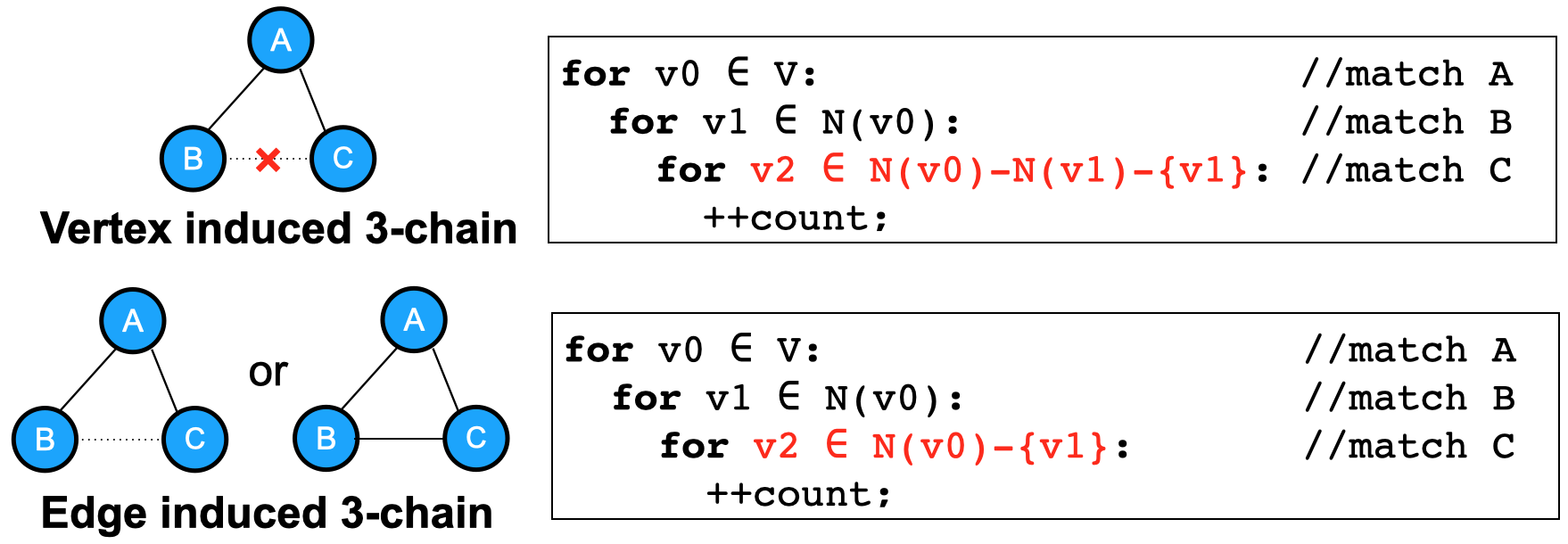}
    \caption{Vertex/edge induced embedding enumeration}
    \label{fig:vertex_edge_enumeration}
\end{figure}

{\bf Vertex vs. edge induced embedding enumeration.} Both kinds of embedding 
can be enumerated with a vertex-set-based method.
Figure~\ref{fig:vertex_edge_enumeration} shows 
the enumeration of 3-chain pattern (A,B,C). 
In a vertex induced 3-chain, B and C cannot be connected,
otherwise (A,B,C) forms a triangle, not a 3-chain.
It can be realized by excluding ``$-N(v1)$''
when matching C: $v2$ cannot
be the neighbor of both A and B.
In an edge induced 3-chain, As long as the edge (A,B)
and (A,C) are included, whether B and C are connected
does not matter: both cases lead to valid 
edge induced 3-chain embeddings. 
It can be realized by removing the constraint ``$-N(v1)$''
for $v2$. 

\proj supports both kinds of pattern enumeration. 
Unless explicitly mentioned,
embedding refers to 
edge induced embedding, which is assumed by 
pattern decomposition algorithm. 
For vertex induced embedding, the pattern count 
({\em cnt}) can be obtained from 
the counts of edge induced embeddings.
In Figure~\ref{fig:vertex_edge_enumeration}, we have the following relation:
\emph{cnt(vertex induced 3-chain)=cnt(edge induced 3-chain)-3$\times$cnt(edge induced triangle)}.
The count of triangle is multiplied by 3 because 
one triangle can lead to three 3-chains by removing 
different edges. 
Our cost model can evaluate the performance
of enumerating vertex induced embedding 
between the two options:
(1) direct method similar to  Figure~\ref{fig:vertex_edge_enumeration}; or 
(2) indirect method by enumerating multiple
edge induced embeddings with pattern decomposition.
It is possible that option (2) is estimated to 
be slower, for which
the compiler falls back to option (1)
without using pattern decomposition.

\begin{figure}[htbp]
    \centering
    \includegraphics[width=\linewidth]{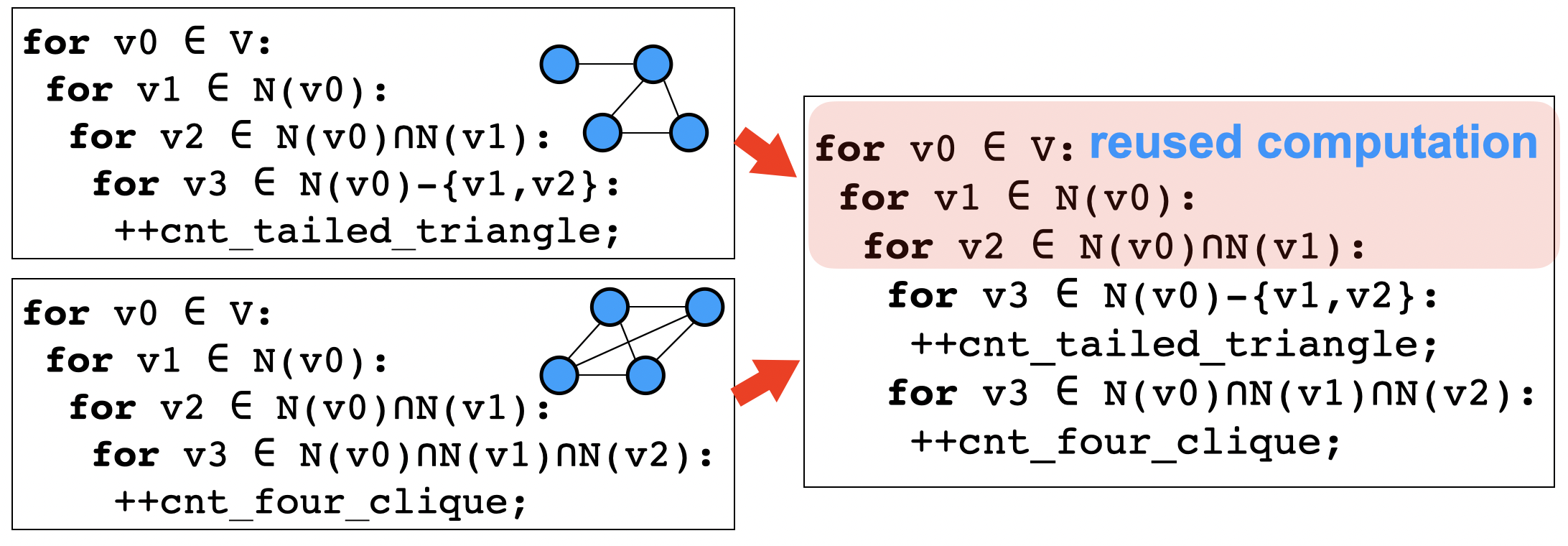}
    \caption{\revision{Computation Reuse}}
    \label{fig:compute_reuse}
\end{figure}

{\bf Optimization 1: symmetry breaking}. This technique ensures that the  
same embedding is only enumerated once  
by enforcing restrictions
on vertices IDs of the embeddings~\cite{grochow2007network,mawhirter2021graphzero,jamshidi2020peregrine}.
In Figure~\ref{fig:gpm_background}, if a subgraph ($v0$,$v1$,$v2$,$v3$) matches pattern vertex 
(A,B,C,D), then 
($v1$,$v0$,$v2$,$v3$), ($v0$,$v1$,$v3$,$v2$), ($v1$,$v0$,$v3$,$v2$)
can also match these pattern vertices---the same embedding will be enumerated for 4 times. 
The redundancy can be eliminated with the restrictions on vertex IDs, e.g., $v0<v1$ and $v2<v3$, so that only one of the matchings is preserved.
The preserved matching is called
automorphism canonical.
For a pattern, 
the set of restrictions can be generated by
by finding equivalent vertices according to pattern automorphisms~\cite{grochow2007network},
which is used in Peregrine~\cite{jamshidi2020peregrine} and GraphZero~\cite{mawhirter2021graphzero}.
There may exist multiple possible sets of restrictions, GraphPi~\cite{shi2020graphpi} utilizes a cost model to select the breaking restrictions that lead
to the best performance.

{\bf Optimization 2: computation reuse}.
It is profitable to schedule the common computations
shared by the same or different patterns together~\cite{mawhirter2019automine}. 
In Figure~\ref{fig:compute_reuse}, the first three loops of the enumeration process of edge-induced 4-cliques and tailed-triangles (left) are the same. The compiler can merge these loops 
so that the common computation is only performed
once (right). Computation reuse can benefit applications
that need to enumerate multiple patterns such as 
Frequent Sugbraph Mining. 
With pattern decomposition, since the algorithm needs
to enumerate subpattern embeddings for even a single 
pattern, this optimization may lead to more benefits. 

\subsection{Prior GPM Systems}

\noindent\textbf{General-purpose GPM systems.}
Arabesque~\cite{teixeira2015arabesque} 
enumerates all subgraphs and relies on expensive
isomorphism checks to identify pattern embeddings. 
RStream~\cite{wang2018rstream} is an out-of-core
GPM system that 
combines streaming processing and relational joins to
optimize for disk I/O accesses.
Peregrine~\cite{jamshidi2020peregrine} utilizes pattern properties (e.g., symmetry) to improve system efficiency.
Pangolin~\cite{chen2020pangolin} and Sandslash~\cite{chen2021sandslash} provide flexible and low-level interfaces so that users can implement more advanced algorithms manually. 
Fractal~\cite{dias2019fractal}, G-miner~\cite{chen2018g} and G-thinker~\cite{yan2020g} are three recent GPM systems focusing on distributed execution.
Automine~\cite{mawhirter2019automine}, GraphZero~\cite{mawhirter2021graphzero} and GraphPi~\cite{shi2020graphpi} are compilation-based systems that generate high-performance GPM programs based on high-level pattern specifications.

\begin{figure}[htbp]
    \centering
    \includegraphics[width=\linewidth]{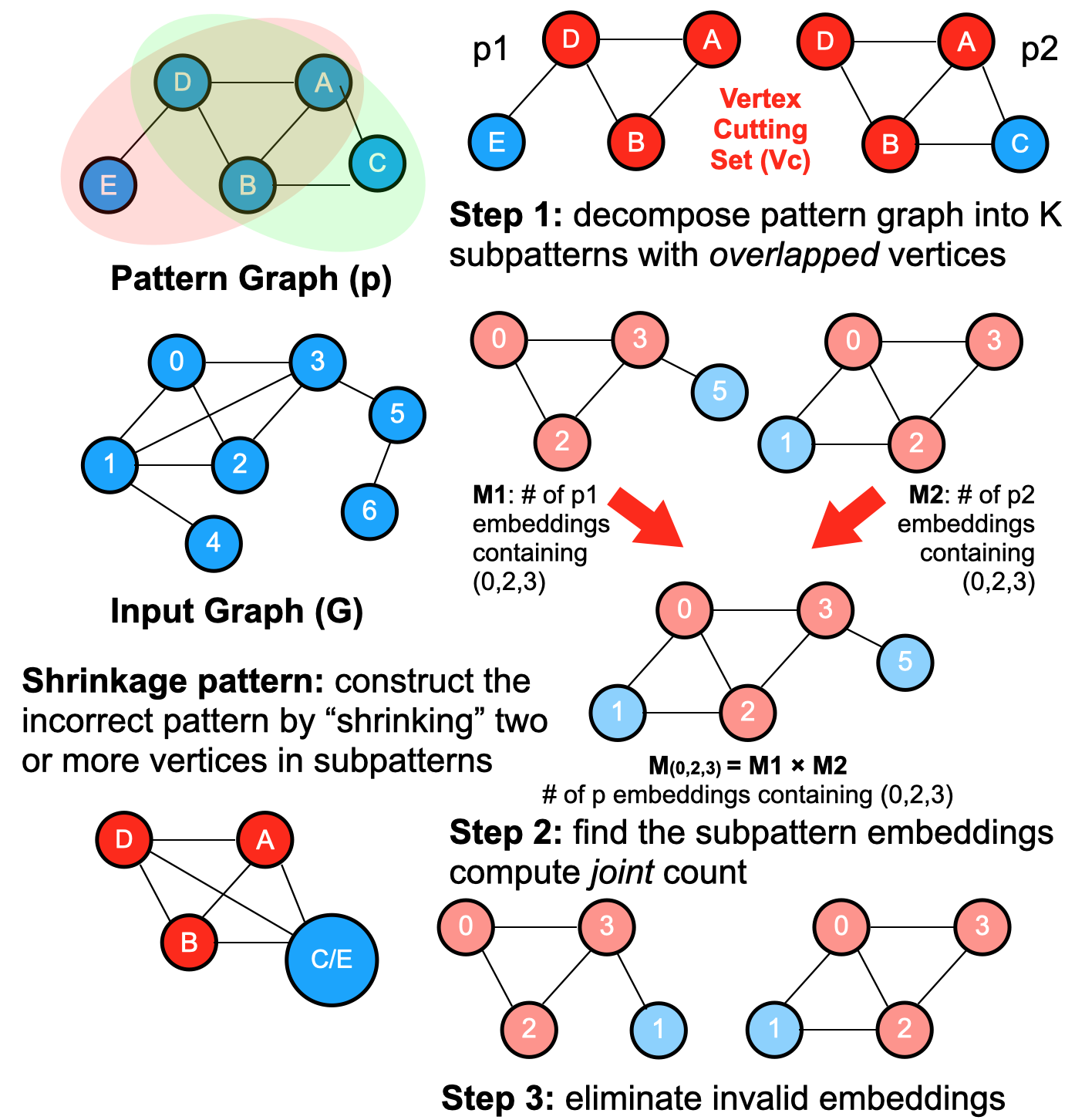}
    \caption{\redasplos{Pattern Decomposition Algorithm Overview}}
    \label{fig:example_pattern_graph}
\end{figure}

\noindent\textbf{Specialized GPM implementations.}
There are a large number of {\em hand-optimized} implementations targeting various GPM applications on different platforms.
OPT~\cite{kim2014opt} is a disk-based triangle solver that discovers all triangle-shape subgraphs.
~\cite{shun2015multicore} is a cache-friendly triangle solver designed for single-node multi-core environments. 
PDTL~\cite{giechaskiel2015pdtl} and DistTC~\cite{hoang2019disttc} are fast distributed triangle discovering implementations.
Frequent subgraph mining (FSM) is another important class of GPM workloads aimed to find frequent 
patterns of interest.
gSpan~\cite{yan2002gspan} is an FSM solver based on DFS.
GRAMI~\cite{elseidy2014grami} avoids redundant subgraph enumeration in FSM to improve performance.
ScaleMine~\cite{abdelhamid2016scalemine} is a distributed FSM solver with a novel load-balancing approach.
DistGraph~\cite{talukder2016distributed} is a distributed FSM system that leverages graph partitioning to support massive datasets.
Other specialized GPM implementations also include those for clique-shape subgraph finding~\cite{danisch2018listing,xiang2013scalable,lu2017finding}, approximate pattern counting~\cite{pagh2012colorful,jha2015path,bressan2018motif,wang2017moss}, and subgraph counting/listing~\cite{pinar2017escape,ahmed2015efficient,ma2012distributed,han2013turboiso,shao2014parallel,lai2015scalable,kim2016dualsim,serafini2017qfrag,bhattarai2019ceci,bi2016efficient,ammar2018distributed}.
These specialized implementations achieve high-performance thanks to various algorithmic optimizations but incur considerable
programming effort. 

\section{\proj System}
\subsection{Pattern Decomposition Method}
\label{sec:decomposition_algorithm}

This section introduces the pattern decomposition
algorithm~\cite{pinar2017escape,ribeiro2019survey}
at a conceptual level.
The highlights of the algorithm with a concrete
pattern graph ($p$) and an input graph ($G$) are shown
in Figure~\ref{fig:example_pattern_graph}.
The algorithm contains three key steps. 

\begin{itemize}[leftmargin=*]
\item {\bf Pattern decomposition.} The original pattern is decomposed into 
subpatterns by choosing a {\em vertex cutting set} ($V_C$), i.e., a subset of pattern graph vertices, 
of which the removal breaks p into several connected components. 
The connected components can be merged with $V_C$ to produce subpatterns. 
Since $V_C$ does not exist in a clique pattern, 
this pattern
cannot benefit from the pattern decomposition. However, clique counting is typically fast and not the performance bottleneck.
The example has two subpatterns 
$p1$: (\textbf{A},\textbf{B},\textbf{D},E) and 
$p2$: (\textbf{A},\textbf{B},C,\textbf{D}) with
the vertex cutting set (A,B,D).

\item {\bf Vertex set enumeration and joint counting. }
This step enumerates the embedding matching
$V_C$ and then 
calculates the number of subpattern embeddings from each 
$V_C$ embedding. 
In the example, $M1$ and $M2$ are the number of 
embeddings matching $p1$ and $p2$, respectively.
They both contain ($v_0$,$v_2$,$v_3$) matching $V_C$.
($v_0$,$v_2$,$v_3$,$v_5$) and ($v_0$,$v_1$,$v_2$,$v_3$)
are two examples of embeddings matching $p1$ and $p2$.
Then the $M1$ and $M2$ subpattern embeddings are conceptually
{\em joined}, leading to $M1 \times M2$ full pattern ($p$) embeddings. 
The embedding of $p$ are not materialized. 

\item {\bf Invalid embedding elimination. }
It identifies and eliminate {\em invalid} embeddings of $p$.
It happens when the join of valid subpattern embeddings 
does not produce a valid embedding of the whole pattern. 
The example shows two embeddings of $p1$ and $p2$ with four vertices 
$v_0$, $v_1$, $v_2$, and $v_3$, but the joint embedding
is not a valid embedding of $p$. 
The reason is that the selection of C and E happen to be the same.
Pinar {\em et al.}~\cite{pinar2017escape} proposes to 
explicitly construct the ``invalid'' patterns, known 
as {\em shrinkage pattern}, by shrinking at least two vertices in $p$ 
in different subpatterns.
We can enumerate the shrinkage patterns, get the count, and 
exclude them from the count calculated in step 2.
The final count of $p$ is further adjusted by dividing 
pattern-dependent multiplicity as previous works~\cite{mawhirter2019automine}. 
\end{itemize}

\subsection{Design Principles}
\label{sec:motiv}

To adopt the algorithm
to a GPM system, we take a compilation-based approach
to automatically generate algorithms for 
arbitrary patterns with different $V_C$ and pattern 
vertex matching orders.
The various implementation candidates are evaluated based
on a cost model so that the one with the best performance can be chosen.
This approach addresses
two key factors affecting performance. 

\begin{itemize}[leftmargin=*]
\item {\bf Incompatible with symmetry breaking.}
Two symmetry embeddings of a subpattern may lead to 
different embeddings of the whole pattern, if only one subpattern
embedding is counted with symmetry breaking, 
the whole pattern embedding will 
be under-counted. 
The compilation-based approach allows a novel pattern-aware loop
rewriting optimization (Section~\ref{sec:loop_rewriting})
that partially recovers the 
performance lost due to the lack of symmetry breaking
in subpatterns. 

\item {\bf No guaranteed performance improvement. }
Some subpatterns after the 
decomposition may be very frequent, which 
can actually
{\em increase} enumeration cost. 
An accurate cost models (Section~\ref{sec:cost_model}) 
can be used to choose the implementations that speedup execution.
The accuracy is crucial for our approach 
since the cost model is required 
to prevent performance degradation.

\end{itemize}

\section{Partial-embedding API}
\label{sec:advanced_app}



\subsection{Motivation}
\label{api_motiv}

\begin{figure}[htbp]
    \centering
    \includegraphics[width=.9\linewidth]{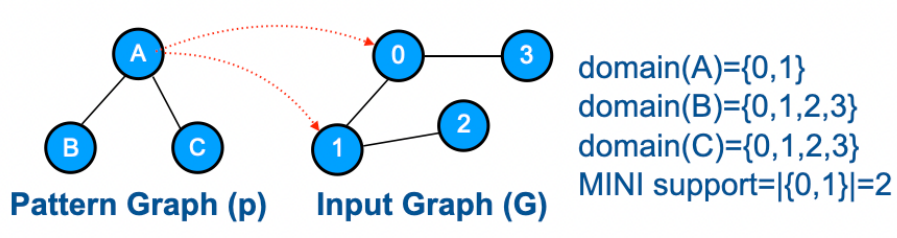}
    \caption{\revision{Building FSM application in a GPM System}}
    \label{fig:fsm_motiv}
\end{figure}

A GPM system typically provides high level API such as 
requesting the system to compute the embedding count of 
a give pattern; or user-defined function (UDF) so that users
can define the operations to be executed on each embedding
identified by the system. 
We use Frequent Subgraph Mining (FSM)~\cite{bringmann2008frequent} as an example to 
illustrate the usage of UDF.
In FSM, \textit{domain}~\cite{jamshidi2020peregrine} of a pattern vertex is the set of input graph vertices that maps to it; 
\textit{MINI support}~\cite{bringmann2008frequent} is the size of the 
smallest domain across all pattern vertices.
FSM discovers \textit{frequent patterns}, whose supports are no less than a user-specified threshold.

Figure~\ref{fig:fsm_motiv} shows the domains and MINI support
of the given pattern graph on an input graph.  
To construct an FSM application, when an embedding of $p$ is 
identified by the system, the UDF updates domain and MINI support
of pattern vertices. 
The problem of using this API with pattern decomposition is, the 
algorithm, which works based on subpattern embedding counts,
is forced to materialize the embedding of the whole pattern 
from a set of subpattern embeddings. 
This procedure will completely diminish the
algorithm advantages. 
For FSM, it is actually 
not necessary because only the {\em mapping} of 
pattern vertex to input vertex needs to be obtained.

\subsection{API Specification}
\label{sec:api_def}

\begin{figure}[htbp]
    \centering
    \includegraphics[width=\linewidth]{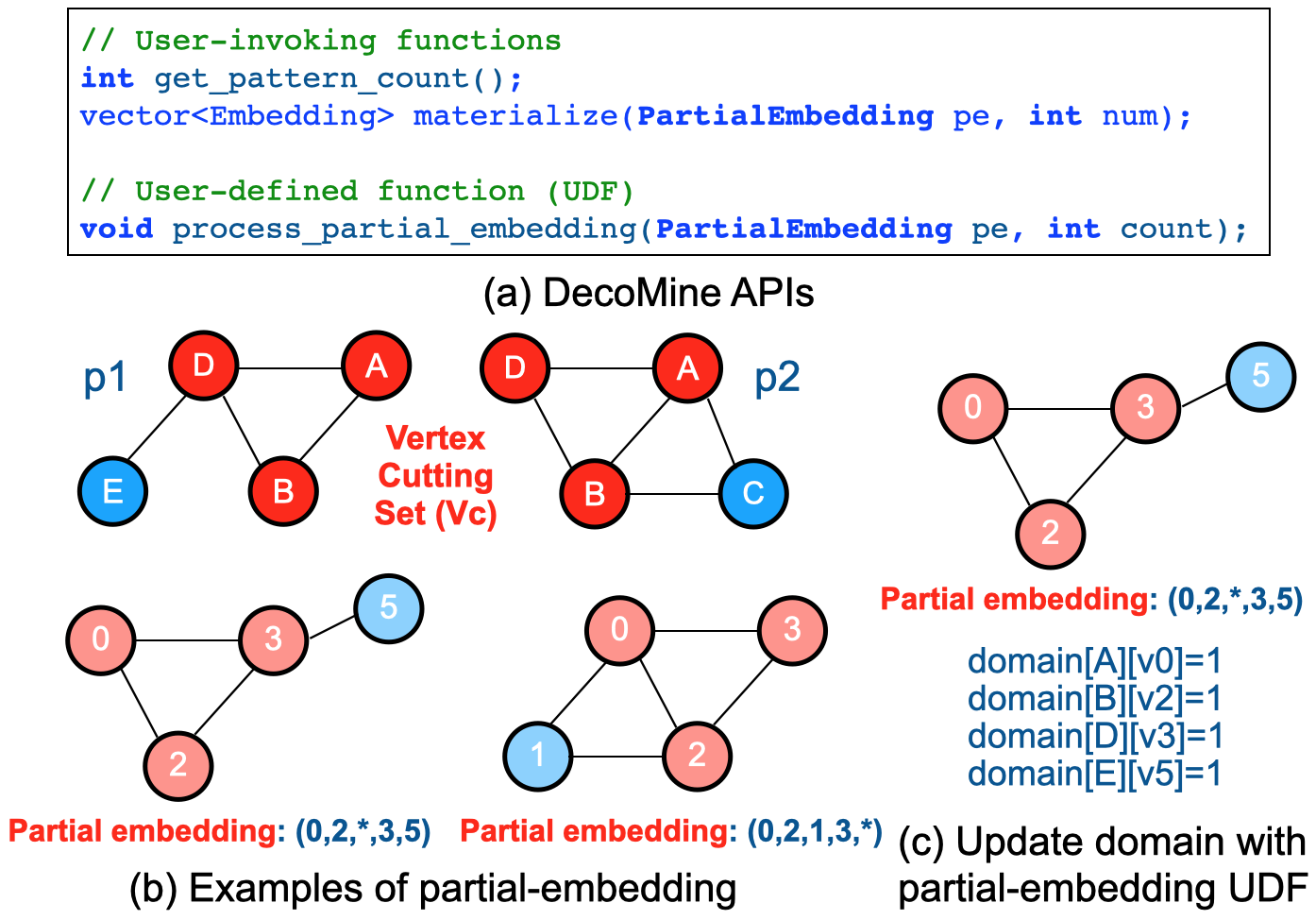}
    \caption{Partial-embedding APIs}
    \label{api_new}
\end{figure}

Figure~\ref{api_new}(a) shows the APIs of \proj.
The key property of the APIs is that they are all 
{\em compatible} with pattern decomposition algorithms
in a sense that they do not lead to unnecessary performance 
overhead, while still being sufficient to build advanced applications. 
The first API \texttt{get\_pattern\_count()} simply returns 
the count of a given pattern. 
Since pattern decomposition algorithm is designed for 
pattern counting, the API does not introduce additional overhead.

The other two APIs are centered around the concept of 
{\em partial embedding}---an embedding of a subpattern. 
The UDF function \texttt{process\_partial\_embedding}
allows the users to define operations on each partial
embedding \texttt{pe} passed from the system. 
The number of whole pattern embeddings that can be 
expanded from \texttt{pe} is also passed in \texttt{count}.
The function \texttt{materialize} allows users
to materialize a certain number (\texttt{num})
of whole pattern embeddings from a partial embedding.
Typically \texttt{materialize} function is called inside
\texttt{process\_partial\_embedding} using the partial embedding
passed from the system. 
Figure~\ref{api_new}(b) shows two examples of partial embedding
corresponding for subpatterns $p1$ and $p2$.
Each partial embedding is represented as a tuple of vertices,
each matches a vertex in pattern graph in alphabetical order. 
The $*$ means the subpattern does not have certain vertex (or vertices).
For FSM,
the domain can be updated in \texttt{process\_partial\_embedding}
for the subpattern vertices, see Figure~\ref{api_new}(c).
But we cannot update the domain of the missing vertex (C in the example). How do we ensure the correctness of the application?

\proj ensures two properties:
\begin{itemize}[leftmargin=*]
\item {\bf Completeness}: if a partial-embedding matching a 
subpattern is passed from the system, then all other partial-embeddings of matching the same subpattern will be also passed;
\item {\bf Coverage}: the set of subpatterns must fully cover all vertices of the pattern graph.
\end{itemize}
For FSM, the completeness property 
ensures the correct calculation of domains, 
while the coverage property that we can obtain the domain 
of all vertices in the pattern graph. 
Note that the coverage property is ensured with 
pattern decomposition algorithm due to the way the subpatterns
are constructed: they share the vertices in $V_C$ and 
cover all pattern graph vertices. 

We claim that the partial-embedding API is an {\em elegant}
solution that works intrinsically with pattern decomposition algorithm while providing intuitive user interface. 
The users only need to construct the applications
based on the concept of subpattern and 
partial-embedding, but \revision{\em are not exposed to algorithmic details 
such as how to decompose a pattern or how to leverage shrinkage patterns}, which 
are the responsibility of \proj compiler. 

\subsection{\revision{Applicability}}
\revision{
With the API, 
\proj is able to benefit applications beyond pattern counting as long as they can be implemented with partial embedding materailization.  
For FSM, by only materializing the fragments of embeddings, whose number can be much smaller than whole embeddings, 
the mapping between pattern and input graph vertices 
are correctly discovered.
Another example of using partial materialization is a graph query like ``listing all types (labels) of vertices that are the centers of size-10 star-shape subgraphs (i.e., with no less than 10 neighbors)''.
Users can discover all center vertices from partial embeddings of a star-shape pattern to record the labels. 
For applications that fundamentally require complete materialization (e.g., writing all embeddings to a disk), 
\proj has to fall back to a non-decomposition-based method, resulting in performance similar to existing systems.
}

\section{Generalized Pattern Decomposition}
\label{aggre}

In this section, we propose a generalized decomposition
algorithm designed for the partial-embedding API.
The algorithm shares the matching process of vertices in $V_C$ among
all subpatterns (the embedding matching $V_C$
is denoted as $e_{C}$),
and perform additional 
on-the-fly enumeration to obtain how many
embeddings of each subpattern 
can be extended from $e_{C}$.
This means that 
$e_1$ and $e_2$ (embeddings of $p_1$ and $p_2$) are constructed in two steps:
first generate $e_C$, and then $e_1'$ and $e_2'$, 
which are obtained by the extension 
from the common $e_{C}$.

\begin{algorithm}[htbp]
    \caption{\blue{\proj Algorithm Template}}
    \label{alg:impl_partial_embedding_model}
    \scriptsize
    \begin{algorithmic}[1]
        \STATE \textbf{Input:} Pattern $p$, Cutting Set $V_C$, Matching orders of all subpatterns
        \STATE Decompose $p$ using $V_C$ to generate $K$ subpatterns, and the shrinkage patterns
        \STATE $pattern\_cnt\gets 0$
        \FORALL{$e_C=(v_0,v_1,\ldots,v_{|V_C|-1})$ matching the cutting set $V_C$}
            \STATE $num\_shrinkages\gets $ \textbf{new} HashTable()
            \STATE clear($num\_shrinkages$)
            \STATE $M\gets 1$
            \FORALL{$i \gets 1...K$}
                \STATE $M_i\gets$ the number of $pe$ extending $e_C$ and matching the $i$-th subpattern
                \STATE $M\gets M \times M_i$
           \ENDFOR
           \STATE $pattern\_cnt\gets pattern\_cnt + M$
           \FORALL{$e$ extending $e_C$ and matching one of the shrinkage patterns}
                \STATE $pattern\_cnt\gets pattern\_cnt -1$
                \FORALL{$i\gets 1...K$}
                    \STATE $pe_i\gets$ extract\_subpattern\_embedding$(e, i)$
                    \STATE $num\_shrinkages[pe_i]\gets num\_shrinkages[pe_i] + 1$
                \ENDFOR
           \ENDFOR
           \FORALL{$i \gets 1...K$}
                \FORALL{$pe$ extending $e_C$ and matching the $i$-th subpattern}
                    \STATE $count\gets M/M_i - num\_shrinkages[pe]$
                    \IF{$count > 0$}
                        \STATE \textbf{process\_partial\_embedding}($pe, count$)
                    \ENDIF
                \ENDFOR
           \ENDFOR
        \ENDFOR
    \end{algorithmic}
\end{algorithm}

Algorithm~\ref{alg:impl_partial_embedding_model}
shows the algorithm template from which 
the compiler can generate concrete pattern decomposition
algorithms for a combination of a vertex cutting set $V_C$
and a matching order.
A {\em matching order} is defined as 
$(o_{vc},o_1,...,o_k,o_{s1},..,o_{sn})$, 
where $o_{vc}$ is the matching order of $V_C$ used in Line 4; each 
$o_i$ determines the order that extends
from a match of $V_C$ to the 
$i$-th subpatterns ($k$ 
in total) used in Line 9; and
each $o_{si}$ determines the order that 
extends to the $j$-th shrinkage pattern ($n$
in total) used in Line 12. 
The template generates
algorithms when application uses 
\texttt{process\_partial\_embedding}.
For pattern counting, the pattern count 
is returned without executing Line \todo{14-21}. 
For each $e_c$, three computation steps 
are performed in an iteration (Line 6-21).

The first
step (Line 7-10) enumerates the embeddings
that can be extended from $e_C$ to match 
the subpatterns using $o_c$, 
and gets a count $M_i$ for 
each subpattern.
Multiplying them together (Line 10) obtains
the number of embeddings $M$ after joining
all partial-embeddings. 
By accumulating all $M$ across different $e_C$ (line 11), 
and deducting 
the redundant and invalid embeddings (line 12),
we can get the correct counts of embeddings of the whole pattern.
Due to space limit, we omit the codes
to eliminate multiplicity.

The second step (Line 12-16) 
constructs the hash table for fast invalid embedding count query.
It enumerates the invalid embeddings $e$
that can be extended from $e_C$
based on each shrinkage pattern $i$ with 
matching order $o_{si}$.
For each invalid embedding $e$, 
\texttt{extract\_subpattern\_embedding}
returns the partial-embedding $pe_i$ of 
subpattern$_i$ contained in $e$.
The same $pe_i$ may be contained
in multiple invalid embeddings, 
so each $pe_i$ should be discounted 
multiple times from the 
number of embeddings of the whole pattern that 
 can be extended from the partial-embeddings 
 of subpattern$_i$.
 The discount number for each $pe_i$ 
 is recorded in the hash table
 $num\_shrinkages$ (line 16).
 
The third step (line 17-21) deducts
 the discount number recorded in the second step
 for each subpattern. 
 $M/M_i$ is the number of embeddings
 that can be extended from a partial-embedding
 of subpattern$_i$ before the deduction.

To reduce the overhead, 
for each hash table entry, we add a 64-bit integer field $entry\_valid$, and maintain another hash-table-wise integer $global\_valid$. 
They are both initialized to 0.
An entry is valid only if $entry\_valid$ is equal to $global\_valid$.  
Once the hash table is cleared, we only increase $global\_valid$ by 1 without modifying any hash table entries. 
If overflow happens, we reinitialize all these $valid$ integers to 0.
Thus, the complexity of clearing (Line 6) is reduced to $O(1)$.
This optimization mainly benefits large cutting sets, which result in 
\redasplos{a large number of clearings}.

\section{\proj Cost Models}
\label{sec:cost_model}

\subsection{Problem, Existing Solution and Improvement}

\begin{figure}[htbp]
    \centering
    \includegraphics[width=\linewidth]{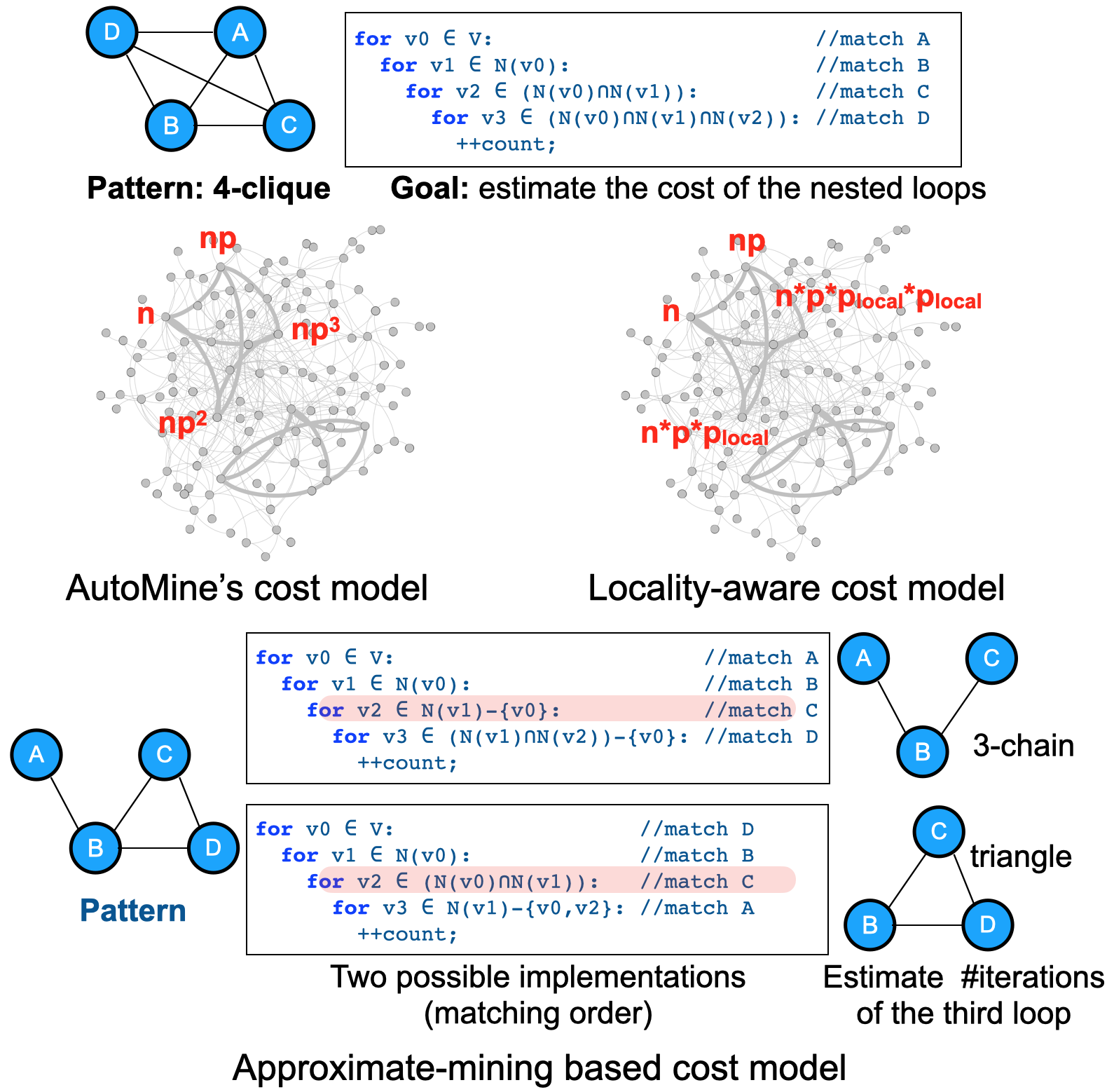}
    \caption{Cost Model Insights}
    \label{cost_model_insights}
\end{figure}

Given a pattern, the pattern enumeration is performed by nested
loops. The problem is to estimate the execution time of the nested
loops, which can be done by estimating the number of iterations
executed in each loop. 
At the top of Figure~\ref{cost_model_insights}, 
we show the nested loops for the 4-clique pattern. 

Automine is the first compilation-based GPM system facing 
the same problem. 
Its cost model assumes \todo{a random input graph}:
the algorithm runs on a random graph with $n$ vertices, where every vertex pair is directly connected by a fixed probability $p$.
For the 4-clique pattern, the number of iterations of the 
$1$-st, $2$-nd, $3$-rd, and $4$-th loop are $n$, $np$, $np^2$, 
and $np^3$, as shown in Figure~\ref{cost_model_insights}.
Unfortunately, the accuracy of the simple cost model is poor.
Consider the livejournal graph with $4.8M$ vertices, 
the connection probability is $1.8 \times 10^{-6}$---the average degree divided by 
the number of vertices.
The estimated cost is $n \times np \times np^2 \times np^3 \approx 1.9 \times 10^{-8}$. However, there are in fact $9.9B$ 4-cliques
in the graph and the actual cost is $(9.9B) \times 4!=238B$.
There is a huge gap between the actual cost and the estimation. 

A simple improvement is to consider
the locality property of real-world graphs. If two vertices are 
$k$-hop neighbors with $k$ less than some threshold $\alpha$,
we increase the probability that they are connected to $p_{local}$ that is much larger than $p$. 
We empirically choose $\alpha = 8$, and 
a $p_{local}$ 
that is close to real-world graphs (e.g., 0.27 for the LiveJournal).
The system allows users to set these parameters.
In Figure~\ref{cost_model_insights}, when estimating the size of $N(v_0)\cap N(v_1)$, since vertices in $N(v_1)$ are 2-hop neighbors of $v_0$, $N(v_0)\cap N(v_1)$ is estimated as $|N(v_1)|p_{local}=npp_{local}$.

\subsection{Approximate-mining Based Cost Model}
\label{sec:approx_cost_model}

To further improve the accuracy of cost prediction, 
we propose to {\em estimate the number of loop iterations at a level
by the approximate count of the corresponding pattern 
reaching that level}.
To illustrate the insights, consider two possible implementations
matching the same pattern in Figure~\ref{cost_model_insights}. 
To estimate the number of iterations for the $3$-rd loop,
the first implementation follows the matching order (A,B,C) while
the second implementation follows the matching order (D,B,C).
Thus, the first three loop levels in the two implementation 
can be estimated by the count of 3-chain and triangle, respectively.

\begin{figure}[htbp]
    \centering
    \includegraphics[width=\linewidth]{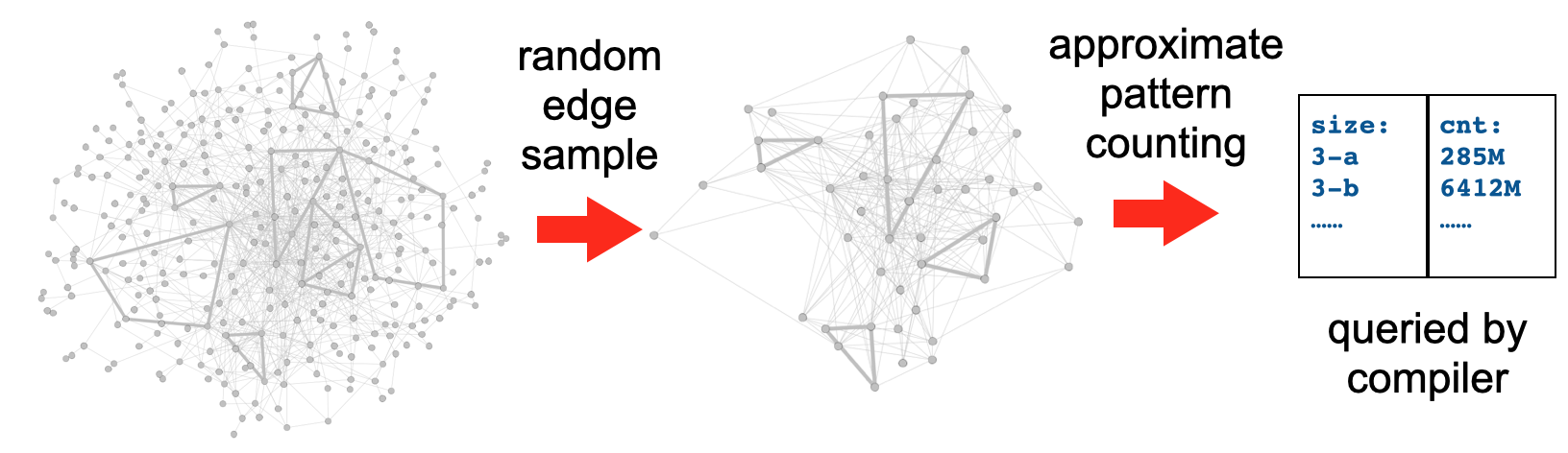}
    \caption{Approximate-mining Cost Model Methodology}
    \label{approx_model}
\end{figure}

Based on this idea, we propose to 
(1) randomly sample a fixed number of edges (e.g., 32M) from input graph; and then 
(2) perform the neighbor sampling~\cite{iyer2018asap} to get the approximate counts of patterns up to a certain size in the sampled
graph. 
These counts are stored in a table that is later
queried by the compiler to get the cost estimation. 
The procedure is shown in Figure~\ref{approx_model}.
The counts in the table are approximate and relative---sufficient for the purpose of cost estimation. 
Unlike vertex sampling, which may drop  
critical structures like hub nodes, 
edge samplings can preserve the hub nodes with high probability. 
Hub nodes are adjacent to a large number of edges, as long as one of them is sampled, the hub vertex will be preserved. 
In practice, collecting approximate counts
for patterns up to 5 vertices is mostly 
enough. \revision{If the counts are missing
due to the large pattern, \proj can quickly run the profiling
on demand and cache the results.}

With a small number of sampled edges, 
the approximate mining algorithms can 
accurately obtain the count of frequent patterns,
while underestimating that of the infrequent ones~\cite{bressan2018motif,bressan2019motivo} since they are less likely to get sampled.
For a cost model, accurately estimating the frequent patterns is 
much more important because they correspond to loops with more iterations, which dominate the execution time.

\subsection{Comparison}

\begin{figure}[htbp]
    \centering
    \includegraphics[width=\linewidth]{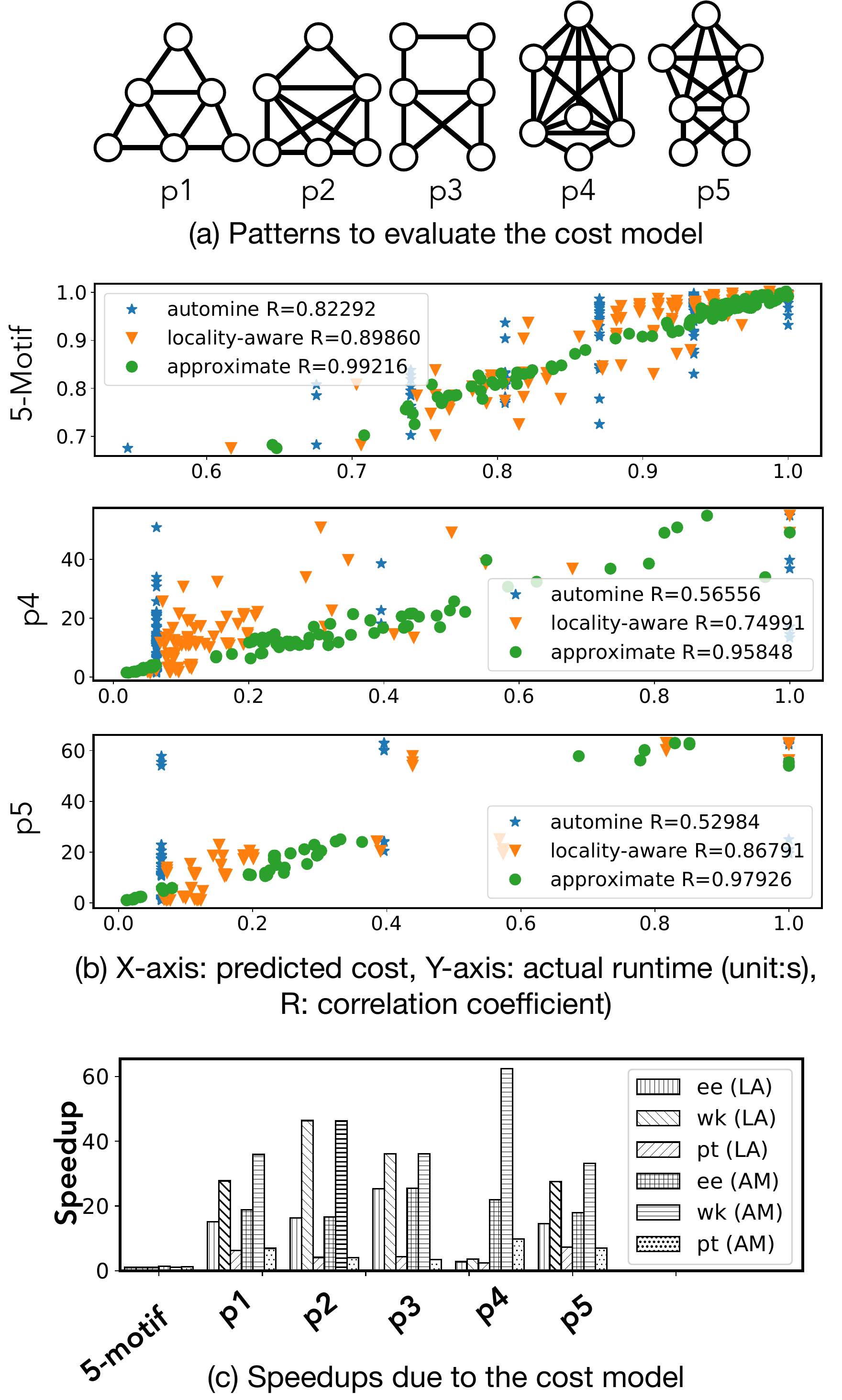}
    \caption{Cost Model Comparison}
    \label{fig:cost_compare}
\end{figure}

We randomly generate 100 implementations
by choosing different cutting sets/matching orders of three applications: 5-motif and two large patterns (p4 and p5) shown in Figure~\ref{fig:cost_compare} (a).
Figure~\ref{fig:cost_compare} (b) shows the relationship between the actual runtime on the EmailEuCore graph~\cite{yin2017local,leskovec2007graph} and the estimated cost by the cost model. 
We use linear correlation coefficiency $R$ (the larger the better) to describe how well the cost model predicts the runtime. 
It shows that the locality-aware model is much better than 
AutoMine's model while approximate-mining based model
provide the highest accuracy. 

The end-to-end performance comparison of different 
cost models shown in Figure~\ref{fig:cost_compare} (c), where 
LA and AM stand for 
locality-aware and approximate mining
cost model, respectively. We see that
the accuracy improvement translates to drastic end-to-end performance improvement---the implementations selected by the locality-aware model and the approximate-mining based model is up to $46.38\times$ and $62.35\times$ (on average $7.12\times$ and $10.92\times$) faster than those selected by Automine's cost model.

The profiling step of approximate-mining based model is very fast, for
CiteSeer (4.5K), MiCo (1.1M), Patents (16.5M), 
LiveJournal (42.9M), and Friendster (1.8B) 
(number of edges in parentheses),
the profiling
only takes 1.96s, 3.50s, 6.64s, 7.14s, and 7.10s.
The benefits of the 
more accurate cost estimation always 
overshadow the small profiling cost.

\section{\proj Compiler}
\label{compiler}

\begin{figure}[htbp]
    \centering
    \includegraphics[width=\linewidth]{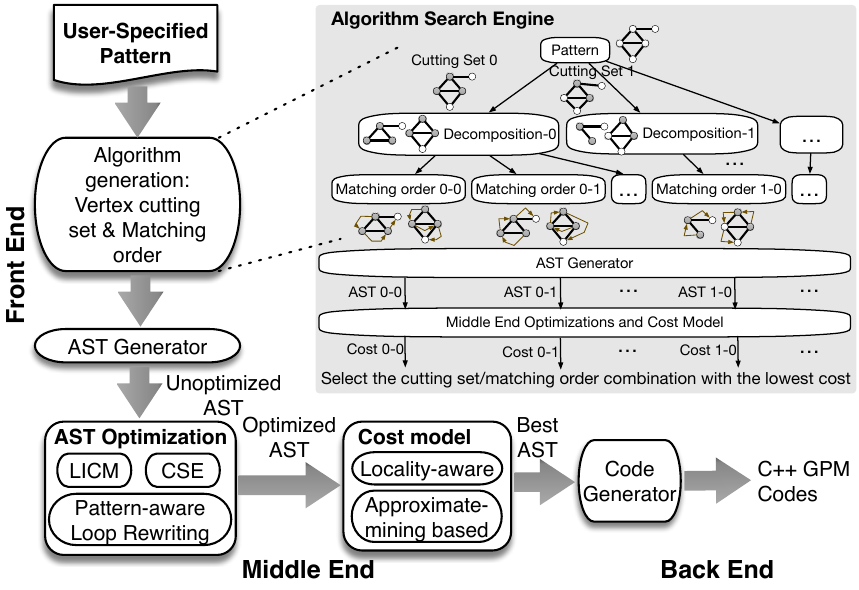}
    \caption{\redasplos{\proj Compiler and Algorithm Search}}
    \label{fig:compiler_workflow}
\end{figure}

The overall compiler workflow is shown in 
Figure~\ref{fig:compiler_workflow}. 
Based on the template in Algorithm~\ref{alg:impl_partial_embedding_model}, the front-end
generates the abstract syntax tree (AST) of the algorithm for each
combination of the vertex cutting set and the vertex
matching order; the middle-end
performs algorithm-oblivious 
optimizations on the AST, whose performance
is estimated by a cost model. 
The back-end generates the C++ GPM codes 
based on the AST with the lowest cost.
The compilation overhead is very low.
For 6-motif, a complicated applications with 112 patterns, compilation takes less than 50ms while running the application on the WikiVote graph (7K vertices) takes 42 minutes.

\subsection{AST Intermediate Representation}
\label{ast_ir}

\begin{figure}[htbp]
    \centering
    \includegraphics[width=\linewidth]{figures/compiler_optimization.pdf}
    \caption{AST-level Optimizations}
    \label{fig:ast_example}
\end{figure}

We use 
the Abstract Syntex Tree (AST) 
as the Intermediate Representation (IR)
to capture the vertex-set-based matching process of subpatterns.
The AST of a vertex-set-based pattern matching process contains five types of nodes. 
1) Loop nodes, which iterate a vertex-ID variable over a vertex set. 
2) Vertex-set operation nodes, which perform an operation to generate a new vertex-set variable. Currently, supported operations include set intersection, subtraction, copy assignment and trimming (eliminating the elements smaller/larger than a lower/upper-bound).
We also support a special vertex-set loading operation that takes a vertex-ID variable as input and loads the corresponding neighbor vertex set.
3) Arithmetic operation nodes, which perform an operation (e.g., addition, multiplication) on scalar variable(s) and produce an output. 
4) Hash table operation nodes, which perform query/update operations to a hash table variable.
5) A virtual root node.
Figure~\ref{fig:ast_example} shows
the AST for a nested loop (a) and two conventional optimizations (b).
Global variables are defined within the root node.
We require that the update to global variables should be associative and commutative,
which is critical to the correctness of our parallelization strategy.
It is satisfied in GPM applications since the ordering of embedding enumeration does not matter.

\subsection{Pattern-aware Loop Rewriting}
\label{sec:loop_rewriting}

For a pattern graph $p$, if 
a subgraph $p'$ is symmetric, 
we can remove the redundant enumeration within
$p'$, while preserving the same
enumeration for $p$ as
no symmetry breaking. 
The key idea is to only enumerate the
subgraphs that match $p'$ once by enforcing
restrictions and {\em compensate} the skipped
computations for $p$.
Figure~\ref{fig:ast_example} (c) shows an example, 
where $p$ contains $p'$, a triangle, which is
symmetric, we can first enumerate the subgraphs
matching the triangle once but
compensate the computation to extend 
matched triangles to embeddings of $p$
for {\em six} times---the multiplicity
of triangle pattern.

Since pattern enumeration is expressed 
as nested loops, this optimization is 
essentially {\em pattern-aware loop
rewriting (PLR)}.
Figure~\ref{fig:ast_example} (c) shows the transformation
based on the example, the extension of 
triangles matched in the first three
loops are expressed as an AST subtree. 
With restrictions, $v_1<v_0$ and 
$v_2<v_1$, the AST subtree is 
immediately scheduled five additional times
for equivalent triangles eliminated
by the restrictions. 
Compared to codes without symmetry 
breaking, PLR improves performance 
for two reasons: 1) eliminate redundant
enumeration for $p'$; and more interestingly,
2) increase the opportunity of CSE when
multiple iterations are scheduled together
in the compensation. 
We can see 2) from Figure~\ref{fig:ast_example} (c), 
when the
\texttt{AST-SUBTREE} are grouped together, 
the paired intersections only need
to perform once. 

When the whole pattern $p$ is symmetric, 
PLR still performs 
strictly more computation than 
applying standard symmetry breaking on $p$.
We attempt to apply PLR to the 
first $k$ loops of 
the vertex cutting set enumeration that 
are shared by all subpatterns.
If PLR is possible, the cost of the optimized 
AST is estimated by the cost model to 
determine: 1) whether PLR is beneficial; 
and 2) the best $k$ value. 
The reason that PLR may lead to worse
performance is that, the benefit of 
eliminating redundant enumeration in $p'$
is overshadowed by code expansion for
compute compensation. 
\revision{Note that there might be symmetry unexploited outside the subpattern $p'$. We leave exploiting such symmetry as our future work.}

\subsection{Algorithm Search}
\label{algo_search}


\proj's algorithm search is performed on optimized ASTs
in a search space determined by 
two algorithm-level decisions:
(1) how to decompose a pattern; and 
(2) the matching order defined in Section~\ref{aggre}.
Generating all decomposition candidates of a pattern with $n$ vertices and $m$ edges consists of two steps.
The first step generates all cutting set candidates by brute force.
All $2^n$ subsets of the $n$ vertices are enumerated, and each of them is checked to see whether it can break the pattern.
This step takes $O(2^n(n+m))$ runtime.
The second step constructs the subpatterns and shrinkage patterns for each cutting set.

\subsection{Code Generation}
\label{code_gen}
The compiler back-end generates
C++ implementations. 
We choose to parallelize the outer-most loop, whose iterations 
are statically divided and assigned to computation threads before execution.
To achieve load balance, 
the compiler uses
fine-grained work stealing to allow idle threads to steal iterations 
that are statically 
assigned but have not been executed by others. 
We use privatization to ensure
thread-safe updates to global variables.
Each thread maintains a private copy for each global variable to accumulate the updates during the execution. 
The copy is updated to the
corresponding global variable using mutex lock. 
The associative and commutative updates to global variables (Section~\ref{ast_ir})
ensures the correctness of the implementation. 
When 
multiple patterns are enumerated concurrently,
the compiler performs
computation reuse optimization 
across different patterns.

\subsection{\revision{Support to Constraints on Labels}}
\revision{
With the generalized pattern decomposition method, 
\proj is able to support label constraints that can be represented by one or multiple sub-constraints, each of which applies to a part of an embedding.
If a constraint $F(e)$ ($e$ is an embedding) can be expressed as 
$F_1(e_1)\land F_2(e_2)\land \ldots \land F_k(e_k)$, 
in which $e_i$ ($1\le i\le k$) is
a fragment of $e$, 
\proj can carefully choose a cutting set so that each sub-constraint can be resolved by only partially materializing $e$. 
For example, a constraint (on the pattern in Figure~\ref{fig:example_pattern_graph}) like ``A, B and C must have different labels and B, D, E must have the same label'' can be supported. 
If $F(e)$ cannot be converted into this representation, or such a cutting set doesn’t exist, the system has to fall back to non-decomposition methods since resolving the constraint depends on all embedding vertices.
}
\section{Experiments}
\label{sec:evaluation}

\subsection{Evaluation Methodology}
\label{sec:eval_method}

\noindent \textbf{System configuration.} 
We conducted experiments on an 8-node cluster. Each node has two 8-core Intel Xeon E5-2630 V3 CPUs and 64GB DDR4 RAM. 
The experiments of Arabesque and Fractal use the whole cluster.
All other systems except for Pangolin-GPU are tested on one node with an additional 2TB NVMe SSD.
Pangolin-GPU uses an NVIDIA V100-32GB GPU. 
Unless specified, the reported runtime is the average of three runs excluding graph loading time.
By default, \proj uses the approximate mining based cost model.

\noindent \textbf{Graph mining applications.} 
\textit{Motif Counting (MC)}~\cite{prvzulj2007biological} aims at counting all connected vertex-induced patterns with a particular size. 
\textit{Frequent Subgraph Mining (FSM)}~\cite{bringmann2008frequent,jiang2013survey,abdelhamid2016scalemine,elseidy2014grami} discovers all frequent labeled patterns in an input graph. 
\textit{Pseudo Clique Mining (PC)}~\cite{uno2010efficient} 
counts vertex-induced pseudo clique patterns of a given size. 
\begin{table}[htbp]
\centering
\scalebox{0.95}{
\begin{tabular}{c|c|c|c|c}
         \hline
         Graph & Abbr. & |V| & |E| & |L| \\
         \hline
         CiteSeer~\cite{nr,geisberger2008better,bader2012graph} & cs & 3.3K & 4.5K & 6 \\
         EmailEuCore~\cite{yin2017local,leskovec2007graph} & ee & 1.0K & 16.1K & 42 \\
         WikiVote~\cite{leskovec2010signed} & wk & 7.1K & 100.8K & N/A \\
         \hline
         MiCo~\cite{elseidy2014grami} & mc & 96.6K & 1.1M & 29 \\
         Patents~\cite{leskovec2005graphs} & pt & 3.8M & 16.5M & N/A \\
         LiveJournal~\cite{backstrom2006group,leskovec2009community} & lj & 4.8M & 42.9M & N/A \\
         \hline
         Friendster~\cite{yang2015defining} & fr & 65.6M & 1.8B & N/A \\
         RMAT-100M\cite{chakrabarti2004r} & rmat & 100M & 1.6B & N/A \\
         \hline
    \end{tabular}
}
    \caption{Graph Datasets~\cite{snapnets}}
    \label{tab:graph_datasets}
\end{table}
A pattern is a pseudo clique if the number of its edges is no less than $n(n-1)/2 - k$, in which $n$ is the number of vertices in the pattern, and $k$ is a pre-defined parameter. All pseudo cliques with $n$ vertices can be obtained by deleting at most $k$ edges from an $n$-clique. In our experiments, we choose $k=1$. 


\noindent \textbf{Graph datasets.} 
Table~\ref{tab:graph_datasets} shows 
the graph datasets. 
The largest dataset is Friendster with roughly 1.8 billion edges. The RMAT-100M graph is synthesized by the RMAT 
generator\cite{chakrabarti2004r,wsvr}
using default parameters.
{We preprocessed all datasets to delete duplicated edges and self-loops.}

\noindent \textbf{In-house Automine implementation.} 
We implemented our 
\begin{table}[htbp]
    \centering
    \centering
    \scalebox{1.}{
    \begin{tabular}{c|c|c|c}
         \hline
         App & Graph & Our Impl. & Original Impl. \\
         \hline
         \multirow{4}{*}{3-MC} & wk & 27.3ms & 34.5ms \\
         & mc & 161ms & 230ms \\
         & pt & 0.9s & 1.9s \\
         & lj & 9.0s & 13.4s\\
         \hline
         \multirow{4}{*}{4-MC} & wk & 7.0s & 11.5s \\
         & mc & 31.7s & 45.2s \\
         & pt & 24.3s & 82.1s \\
         & lj & 457m & 367m \\
         \hline
         \multirow{3}{*}{5-MC} & wk & 4345s & 5300s \\
         & mc & 2.91h & 5.56h \\
         & pt & 54m & 117m \\
         \hline
    \end{tabular}
    }
    \caption{In-house Automine vs. Automine Runtime in ~\cite{mawhirter2021graphzero}}
    \label{tab:compare_automine}
\end{table}
own Automine (AutoMineInHouse) with all optimizations in ~\cite{mawhirter2019automine}. 
Table~\ref{tab:compare_automine} compares
the performance of 
AutoMine reported in \cite{mawhirter2021graphzero}
and AutoMineInHouse using a similar machine. 
AutoMineInHouse 
is faster in almost all cases except for the 4-motif counting on the lj graph, which may be due to the minor hardware discrepancy. 

\subsection{Overall Performance}
\label{sec:eval_overall}

\begin{table}[htbp]
    \centering
    \scalebox{0.7}{
    \begin{tabular}{c|c|c|c|c|c}
        \hline
        App. & G & \proj & AutomineInHouse & RStream & Arabesque (8-node) \\
        \hline
        \multirow{6}{*}{\begin{sideways}3-MC\end{sideways}} & cs & 0.14ms & 0.16ms (1.1x) & 142ms (1,014x) & 10.1s (\textbf{72,143x}) \\
        & ee & 0.87ms & 7.3ms (8.4x) & 21.0s (24,138x) & 10.2s (11,724x) \\
        & wk & 7ms & 27.3ms (3.9x) & 17.9m (153,428x) & 12.1s (1,729x) \\
        & pt & 332ms & 931ms (2.8x) & 104.1m (18,813x) & 96.4s (290x) \\
        & mc & 48ms & 161ms (3.4x) & 144.8m (181,000x) & 21.1s (440x) \\
        & lj & 2.7s & 9.0s (3.3x) & T & 24.3m (540x) \\
        \hline
        \multirow{6}{*}{\begin{sideways}4-MC\end{sideways}} & cs & 0.17ms & 4.8ms (28x) & 3.7s (21,765x) & 9.9s (58,235x) \\
        & ee & 9ms & 920ms (102x) & 132.4m (\textbf{882,667x}) & 19.1s (2,122x) \\
        & wk & 60ms & 7.0s (117x) & T & 402.2s (6,703x) \\
        & pt & 1.5s & 24.3s (16x) & T & 68.3m (2,732x) \\
        & mc & 1.3s & 31.7s (24x) & T & 42.8m (1,975x) \\
        & lj & 33.1s & 456.5m (\textbf{827x}) & T & C \\
        \hline
        \multirow{6}{*}{\begin{sideways}5-MC\end{sideways}} & cs & 2.1ms & 332ms (158x) & 146.4s (69,714x) & 11.4s (5,429x) \\
        & ee & 416ms & 104.8s (252x) & T & 19.4m (2,798x) \\
        & wk & 8.1s & 72.4m (536x) & T & C  \\
        & pt & 36.9s & 53.9m (88x) & T & C \\
        & mc & 111.6s & 174.6m (94x) & T & C \\
        & lj & 167.7m & T  & T & C  \\
        \hline
        \multirow{4}{*}{\begin{sideways}6-MC\end{sideways}} & cs & 270ms & 35.9s (133x) & 108.7m (24,156x) & 48.7s (180x) \\
        & ee & 106.6s & 259.0m (146x) & T  & C \\
        & wk & 42.2m & T & T  & C \\
        & pt & 63.0m & T & T  & C  \\
        \hline
        \multirow{4}{*}{\begin{sideways}7-PC\end{sideways}} & cs & 0.3ms & 0.5ms (1.7x) \\
        & ee & 719ms & 67.1s (93x) \\
        & wk & 735ms & 90.8s (24x) \\ 
        & pt & 499ms & 15.7s (31x) \\ 
        \hline
        \multirow{4}{*}{\begin{sideways}8-PC\end{sideways}} & cs & 0.3ms & 0.5ms (1.7x) \\
        & ee & 1.3s & 433.1s (322x) \\
        & wk & 1.2s & 463.0s (387x) \\
        & pt  & 582ms & 86.2s (148x) \\
        \hline
        \multirow{3}{*}{\begin{sideways}FSM-300\end{sideways}} & cs & 2.6ms & 7.7ms (3.0x) & 522ms (201x) & 10.3s (3,962x) \\
        & ee & 0.3ms & 0.3ms (1.0x) & 3.6s (12,000x) & 9.6s (32,000x) \\
        & mc & 210.8s & 242.7s (1.2x) & 149.1m (42x) & C  \\
        \hline
        \multirow{3}{*}{\begin{sideways}FSM-3K\end{sideways}} & cs & 0.3ms & 0.3ms (1.0x) & 77.9ms (260x) & 9.6s (32,000x) \\
        & ee & 0.3ms & 0.3ms (1.0x) & 210ms (700x) & 9.8s (32,667x) \\
        & mc & 513ms & 30.0s (58x) & 141.9m (16,596x) & 157.9s (308x) \\
        \hline
    \end{tabular}}
    \caption{\red{Comparing with Automine/RStream/Arabesque. T: Timeout (12 h) C: Crashed (out of memory/disk space)}}
    \label{tab:overall_ARA}
\end{table}

Table~\ref{tab:overall_ARA} compares 
\proj to AutoMineInHouse, 
RStream~\cite{wang2018rstream}, and 
Arabesque~\cite{teixeira2015arabesque}. 
\revision{The runtimes exclude graph loading and profiling time as they can be amortized with multiple applications.}
{In both $k$-MC and $k$-PC, $k$ refers to the pattern size.}
\red{In FSM, the support thresholds are 300 and 3000. Since FSM requires the labeled input graph, we only evaluate it on CiteSeer, EmailEuCore, and MiCo.}
\redasplos{Similar to previous works~\cite{jamshidi2020peregrine,chen2020pangolin}, we only discover frequent patterns with less than four edges.}
We do not run the Pseudo-Clique Counting experiments for RStream and Arabesque due to the lack of reference implementations. 
\proj outperforms other systems
significantly, achieving speedups 
of up to \maxspeedupautomine, \maxspeeduprstream, and \maxspeeduparabesque over 
Automine, RStream, and Arabesque, 
respectively. 
\revision{
An important reason why existing systems like RStream and Arabesque are so slow is that they have to materialize all embeddings,
which is usually not required by applications.
For example, MC/PC only need the counts of patterns while FSM only needs the pattern vertex domains. 
In contrast, the flexibility of \proj allows users to match the need of applications more precisely and thus avoid unnecessary materialization for higher performance.
}

For MC and PC, on the same graph, for larger pattern sizes, \proj can 
achieve higher speedups over Automine and RStream.
\blue{Compared to Arabesque, the speedup is higher when the pattern size is small due to Arabesque's startup overhead.
When the pattern size is larger, 
this performance issue is mitigated since 
the overhead is amortized and thus the speedup over Arabesque decreases. }

For FSM,
the performance of \proj and Automine 
on the cs (support=3K) and ee graphs
is similar. 
\red{Both datasets are tiny graphs. 
Hence, almost all labeled patterns are filtered away even with a relatively low support threshold, i.e., 3K. As a result, only trivial startup computation overhead is left, leading to a similar performance of Automine and \proj.} 

We also compare \proj with Peregrine~\cite{jamshidi2020peregrine}, Pangolin~\cite{chen2020pangolin} and Fractal~\cite{dias2019fractal}.
We run the four systems for MC and FSM.
\red{We evaluate MC on cs/pt/mc and FSM on mc/lj (lj with randomly synthesized labels).}
As shown in Table~\ref{tab:overall_peregrine}, \proj is consistently faster, and achieves up to \maxspeedupperegrine, \maxspeeduppangolin and \maxspeedupfractal speedup over Peregrine, Pangolin-CPU and Fractal, respectively.
Pangolin ran out of memory and crashed for many benchmarks due to its memory-demanding BFS exploration.
Pangolin-GPU's performance is competitive with \proj.
Nevertheless, it runs on a significantly more expensive device (NVIDIA V100-32GB).
It is around 8K USD, 6.15$\times$ costly compared with the CPUs used by \proj (two Intel Xeon E5-2630 v3, 1.3K USD).
More importantly, Pangolin-GPU fail to run on large graphs/patterns due to the GPU memory limitation (marked as "C").


\begin{table}[htbp]
    \centering
    \scalebox{0.67}{
    \begin{tabular}{c|c|c|c|c|c}
        \hline
        App. & G & \proj & Peregrine & Pangolin(CPU/GPU) & Fractal (8-node) \\ 
        \hline
        \multirow{3}{*}{3-MC} & cs & 0.14ms & 5.8ms & 5.0ms / 0.1ms & 5.9s \textbf{(42,143x)} \\
        & pt & 332ms & 1.4s & 1.4s / 0.2s & 79.7s \\
        & mc & 48ms & 60ms & 280ms / 14.1ms & 12.9s \\
        \hline
        \multirow{3}{*}{4-MC} & cs & 0.17ms & 21.2ms & 15.3ms / 0.7ms & 6.0s \\
        & pt & 1.5s & 11.2s & 329.5s / 8.0s & 141.6s \\
        & mc & 1.3s & 5.3s & 242.7s / 3.7s & 58.4s \\
        \hline
        \multirow{3}{*}{5-MC} & cs & 2.1ms & 41.7ms & 688.3ms \textbf{(328x)} / 1.3ms & 6.1s \\
        & pt & 36.9s & 513.6s & C / C & 4517.0s \\
        & mc & 111.6s & 5,635.1s & C / C & 1240.0s \\
        \hline
        \multirow{2}{*}{6-MC} & cs & 270ms & 0.8s & 14.9s / C & 4.6s \\
        & pt & 63.0m & T & C / C & T \\
        \hline
        FSM-300 & \multirow{3}{*}{mc} & 210.8s & C & C / C & 280.2s \\
        FSM-1K & & 3.1s & 1,782.2s \textbf{(575x)} & C / C & 169.1s \\
        FSM-3K & & 513ms & 189.3s & C / C & 109.4s \\
        \hline
        FSM-1.0M & \multirow{3}{*}{lj} & 27.8s & T & C / C & T \\
        FSM-1.5M & & 27.8s & T & C / C & 3.4h \\
        FSM-2.0M & & 27.8s & T & C / C & 3.3h \\
        \hline
    \end{tabular}    }
    \caption{\red{\proj vs. Peregrine and Pangolin}}
    \label{tab:overall_peregrine}
\end{table}

\begin{figure}[htbp]
    \centering
    \includegraphics[width=\linewidth]{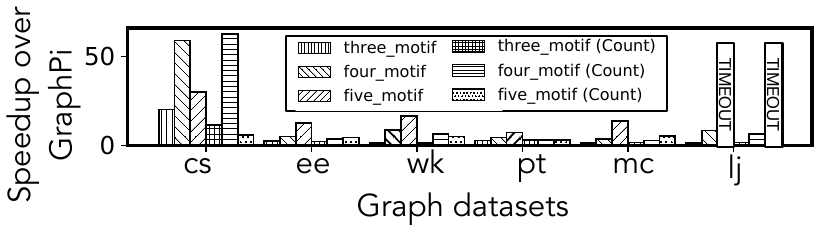}
    \caption{\blue{Comparing with GraphPi}}
    \label{fig:compare_with_graph_pi}
\end{figure}

We compare \proj with GraphPi~\cite{shi2020graphpi} in Figure~\ref{fig:compare_with_graph_pi} for motif mining applications. 
We do not use FSM since GraphPi does not supported it.
The version marked with ``(count)'' indicates
that the pattern counting mathematical
optimization of GraphPi is enabled,
which quickly evaluates the number of iterations of some certain loops.
Our system consistently outperforms GraphPi, and achieves up to \maxspeedupgraphpi speedup.
GraphPi's mathematical optimization significantly improves its performance but is only suitable for pattern counting. 




\noindent\textbf{Comparing with the efficient native algorithm.} We compare 
\proj with ESCAPE~\cite{pinar2017escape} in Table~\ref{tab:comp_native_algorithms}, the fastest expert-tailored
single-thread
decomposition-based 4/5-motif 
\begin{table}[htbp]
    \centering
    \scalebox{1.}{
    \begin{tabular}{c|c|c|c|c|c}
        \hline
        App. & Graph & \multicolumn{2}{c|}{\proj} & GraphPi & ESCAPE \\ 
        \hline
        \#Cores & & 16 & 1 & 1 & 1 \\
        \hline
        \multirow{3}{*}{4-MC} & ee & 9ms & 95ms & 397ms & 32ms \\ 
        & wk & 60ms & 879ms & 5.8s & 312ms \\ 
        & pt & 1.5s & 19.9s & 62.4s & 10.3s \\ 
        \hline
        \multirow{3}{*}{5-MC} & ee & 416ms & 5.4s & 26.5s & 889ms \\ 
        & wk & 8.1s & 121.0s & 617.2s & 12.7s \\ 
        & pt & 36.9s & 557.5s & 1719.5s & 133.9s \\ 
        \hline
    \end{tabular}    }
    \caption{\blue{\proj vs. Native Algorithm}}
    \label{tab:comp_native_algorithms}
\end{table}
counting implementation~\cite{ribeiro2019survey}
with
\redasplos{pattern-specific optimizations}.
With one thread, \proj is on average $4.0\times$ slower than ESCAPE, while GraphPi, even with their mathematical optimization that only applies to pattern counting, is on average $17.3\times$ slower (up to $48.6\times$).
With multicores,  
\proj runs faster than ESCAPE.
\revision{Note that ESCAPE still outperforms \proj in single-thread performance because it adopts an algorithmic optimization that converts each size-5 pattern to a DAG to reduce the exploration space. }

\subsection{Pattern-aware Loop Rewriting}
\label{src:eval_SB}

\begin{figure}[htbp]
    \centering
    \includegraphics[width=\linewidth]{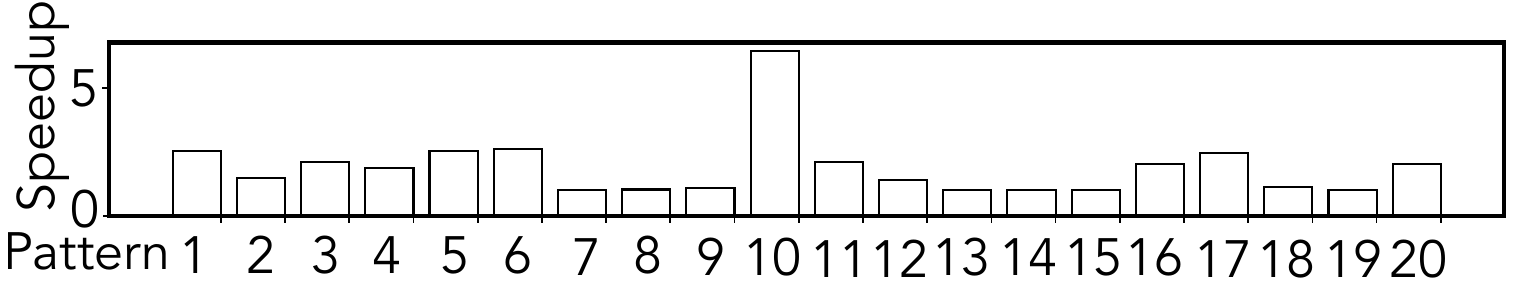}
    \caption{\blue{Speedups due to PLR}}
    \label{fig:analyze_loop_rewriting}
\end{figure}
We generate the counting application for each size-5 pattern except for the 5-clique (since 5-clique cannot benefit from pattern decomposition) (in total 20 patterns) with and without PLR enabled, 
and run them on the Patents graph.
Figure~\ref{fig:analyze_loop_rewriting} shows the speedup brought about by PLR.
The optimization improves the performance by up to $6.5\times$, and benefits more than a half of size-5 patterns.

\subsection{Scalability}
\label{sec:eval_large}


\begin{table}[htbp]
    \centering
    \scalebox{1.}{
    \begin{tabular}{c|c|c|c|c}
         \hline
         G. & |V| & |E| & System & Runtime \\
         \hline
         \multirow{3}{*}{fr} & \multirow{3}{*}{65.6M} & \multirow{3}{*}{1.8B} &  \proj & 1.4h \\
         & & & Peregrine & 29.1h \\
         & & & GraphPi & 15.4h \\
         \hline
         \multirow{3}{*}{rmat} & \multirow{3}{*}{100M} & \multirow{3}{*}{1.6B} & \proj & 1.7h \\
         & & & Peregrine & 39.7h \\
         & & & GraphPi & 10.2h \\
         \hline
    \end{tabular}
    }
    \caption{Large Graphs}
    \label{tab:runtime_larger_graphs}
\end{table}

We evaluate \proj's capability to scale to large graphs by 
comparing it with 
Peregrine and GraphPi, for 4-motif mining on fr and rmat, both with more than one billion edges.
For GraphPi, the pattern counting mathematical optimization is enabled to achieve all its performance benefits. 
We report the runtimes in Table~\ref{tab:runtime_larger_graphs}.
For large workloads like motif mining on large graphs, 
\proj can significantly reduce execution time 
from tens of hours in recent systems to less
than two hours.



\setlength{\intextsep}{2pt}%

\begin{table}[htbp]
    \centering
    \scalebox{1.}{
    \begin{tabular}{c|c|c|c|c}
         \hline
         G. & App. & \proj & Peregrine & GraphPi \\
         \hline
         \multirow{3}{*}{ee} & 6-cycle & 3.4s & 102.7s & 64.8s \\
         & 7-cycle & 249.4s & 6131.9s & 3,674.7s \\
         & 8-cycle & 5.7h & 5.6d & 2.8d \\
         \hline
         \multirow{2}{*}{pt} & 6-cycle & 370.2s & 6913.9s & 1960.0s \\
         & 7-cycle & 4.4h & 2.9d & 23.3h \\
         \hline
         \multirow{2}{*}{wk} & 6-cycle & 136.2s & 5754.9s & 3,248.6s \\
         & 7-cycle & 4.8h & >1 week & 4.0d \\
         \hline
    \end{tabular}
    }
    \caption{Large Patterns (d: days)}
    \label{tab:eval_larger_patterns}
\end{table}

We use cycle mining on ee, pt and wk to evaluate the scalability of \proj to large patterns. 
$k$-cycle mining counts the number of size-$k$ cycles of the input graph.
We keep increasing the pattern size (i.e., $k$) until our system cannot finished it within 24 hours.
The runtime results are reported in Table~\ref{tab:eval_larger_patterns}.
For comparison purposes, we also report the execution times of Peregrine and GraphPi for the same workloads. 
The results show that 
\proj can easily mine large patterns, e.g., $8$-cycle on the ee graph, within a few hours while the other two
systems usually takes {\em days} to complete.
For 7-cycle on wk, \proj reduces the execution time 
{\em from more than one weeks on Peregrine to less than five hours}.


\begin{figure}[htbp]
    \centering
    \includegraphics[width=\linewidth]{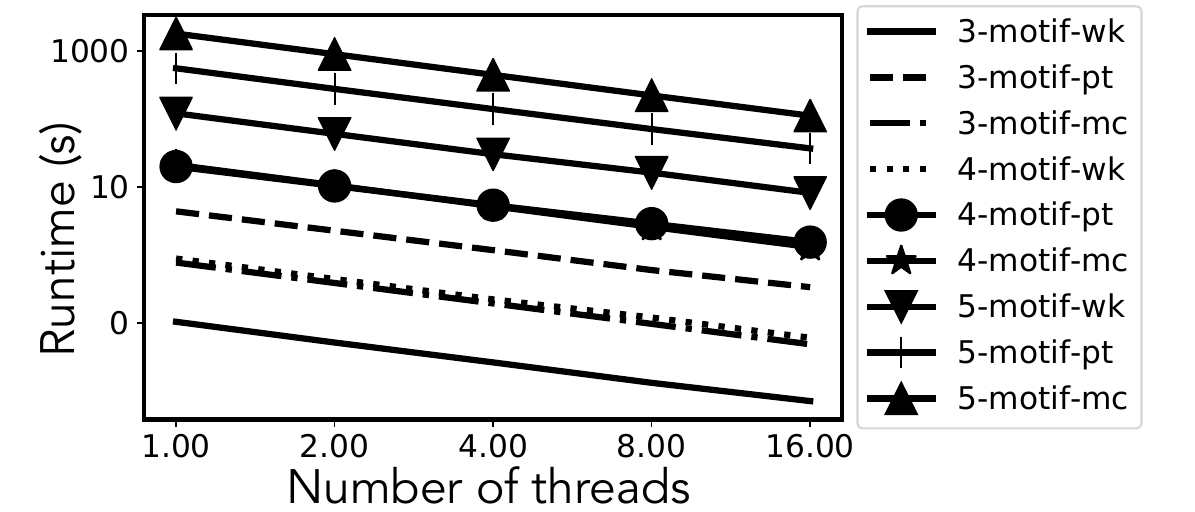}
    \caption{Scalability with MT}
    \label{fig:scalability}
\end{figure}
The scalability of \proj with multi-thread
is shown in Figure~\ref{fig:scalability}. 
\proj achieves almost linear scalability. {The single-thread and 16-thread runtimes of \proj for 5-Motif on Patents are 557.5 seconds and 36.9 seconds, respectively---a speedup of $15.11\times$.}

\subsection{FSM with Various Support Thresholds}
\label{sec:eval_fsm}

\begin{figure}[htbp]
    \centering
    \includegraphics[width=\linewidth]{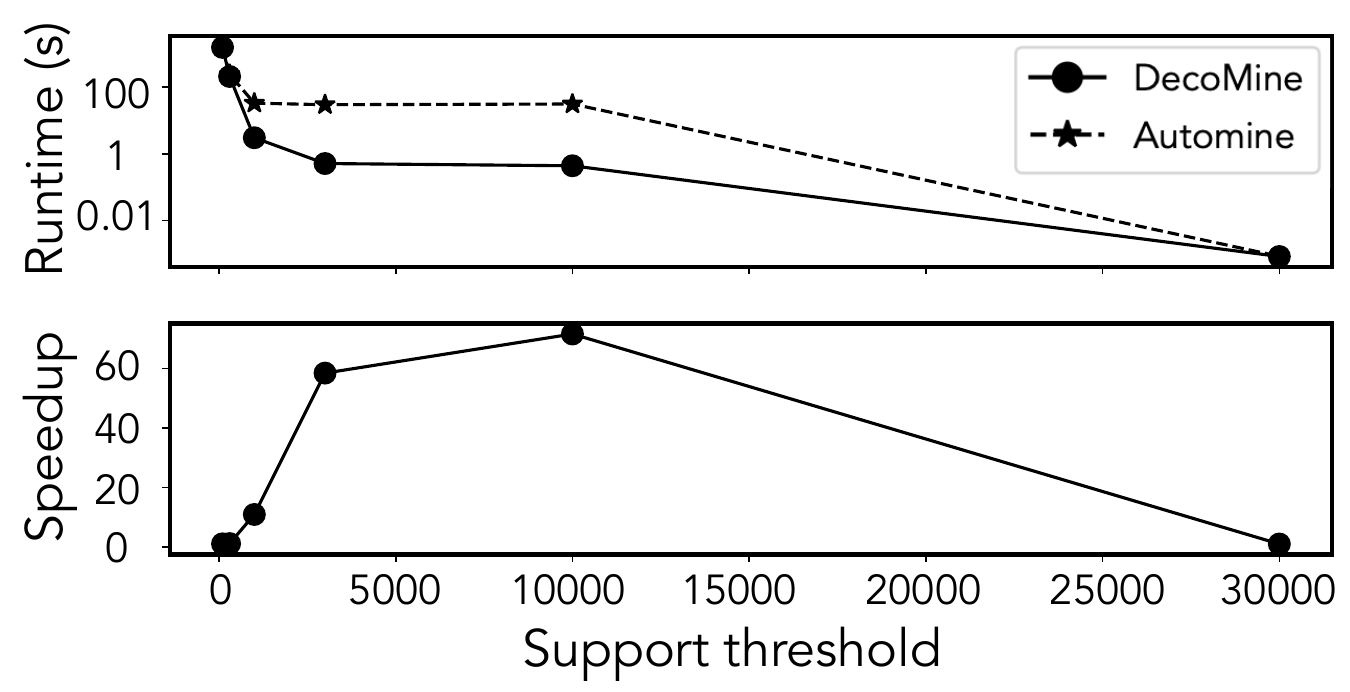}
    \caption{FSM Sensitivity}
    \label{fig:larger_fsm}
\end{figure}
We run \proj and AutomineInHouse for FSM with various support thresholds ranging from 100 to 30K on the MiCo graph to analyze the performance sensitivity with respect to thresholds.
The runtimes and \proj's speedups over AutomineInHouse are reported in Figure~\ref{fig:larger_fsm}.
\proj is consistently faster than AutoMineInHouse in all settings.
We also see that the speedups are small with extremely large and 
small thresholds, while the a peak speedup of roughly $70\times$ is reached when the threshold is 10K.
For extremely large thresholds like 30K, 
almost all patterns are filtered away by the exceedingly high threshold, 
and hence only trivial computation cost is left, leading to the similar performance of \proj and AutomineInHouse.
On the other side, when the threshold is very small (e.g., 100), there are a huge number of lightweight labeled patterns to be processed and hence FSM is mainly bottlenecked by the per-pattern overhead (e.g., clearing data structures, launching computation threads), which cannot be accelerated by pattern decomposition. 

\subsection{\revision{Workloads with Label Constraints}}

\revision{To demonstrate \proj's ability to handle workloads with complicated label constraints, we evaluate a graph query ``counting the subgraphs matching pattern p in Figure~\ref{fig:example_pattern_graph}, in which vertices matching A, B, C must have different labels and vertices matching B, D, E must have the same label'' on \proj and Peregrine.
Thanks to \proj's ability to resolve constraints on partially-materialized embeddings, the system is able to achieve significantly higher performance than Peregrine: for (cs,ee,mc,lj) graphs, the runtimes of \proj are 
(0.35ms,43ms,11.9s,288.4s), while the runtimes of Peregrine are
(2.2ms,975ms,2030.9s,>12h).
}

\subsection{\revision{Compilation Cost}}

\begin{figure}[htbp]
    \centering
    \includegraphics[width=\linewidth]{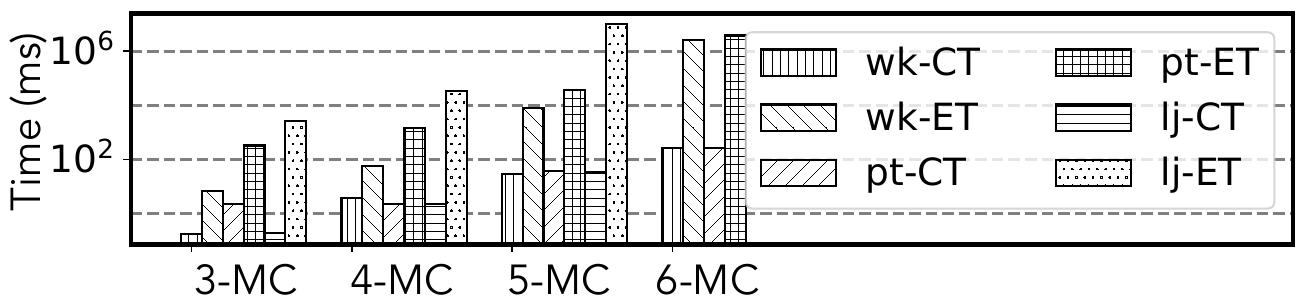}
    \caption{\revision{Compilation Time (CT) vs. Execution Time (ET)}}
    \label{fig:compilation_cost}
\end{figure}

\revision{
The compilation overhead of \proj is negligible compared to the execution time. We compare the compilation and execution time for 3/4/5/6-MC on wk, pt, and lj in Figure~\ref{fig:compilation_cost}. 
The trends on other workloads are similar and we omit them due to space limit. 
The compilation times are {\em orders of magnitude smaller} than execution times. 
Even for 6-MC, the most complicated workloads evaluated in this paper (with 112 size-6 patterns), the compilation takes less than 300ms while the execution on a median-size graph (e.g., wk) takes tens of minutes.
}

\subsection{\revision{Detailed Cost Model Analysis}}

\begin{figure}[htbp]
    \centering
    \includegraphics[width=\linewidth]{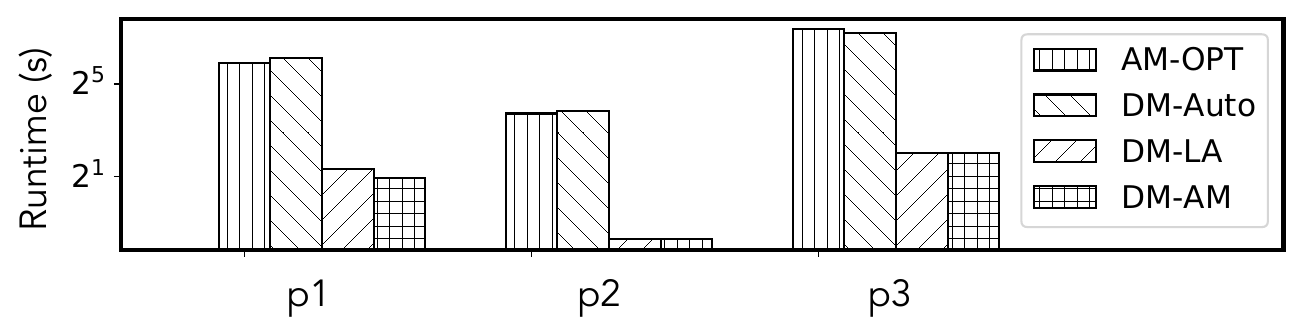}
    \caption{\revision{Comparing \proj with Automine with an Optimal Cost Model (wk graph)}}
    \label{fig:detailed_cost_model}
\end{figure}

\revision{
To analyze the contribution of the improved cost models in a pattern decomposition based system, 
in Figure~\ref{fig:detailed_cost_model},
we compare the runtimes of Automine with a perfect cost model (AM-OPT) (obtained by trying all matching orders on evaluated graphs) and \proj with Automine's cost model (DM-Auto), the locality-aware model (DM-LA) and the approximate mining based model (DM-AM) for patterns p1, p2 and p3 in Figure~\ref{fig:cost_compare} (a).
There are two key observations.
First, even with an ideally perfect cost model, Automine (AM-OPT) is significantly slower than \proj with an improved cost model (DM-LA/AM).
It confirms that algorithmic advantage is the fundamental reason for \proj's high performance.
Second, 
with an inaccurate cost model (DM-Auto),
\proj can be indeed slower than a system without decomposition (AM-OPT) because a bad cutting set is selected.
It confirms the importance of an accurate cost model in \proj.
}

\section{Conclusion}
\label{sec:conc}

This paper proposes a new compilation-based GPM system that 
adopts efficient pattern decomposition algorithms. 
The compiler generates algorithms with 
conventional and GPM-specific optimizations
for different decomposition choices, which
are evaluated based on an accurate cost model.
The executable of the GPM task is obtained from the 
algorithm with the best performance. 
We propose a novel partial-embedding API that 
is sufficient to construct advanced GPM applications 
while preserving pattern
decomposition algorithm advantages. 
The key insight is that the system does not need to 
materialize the embedding of the whole pattern while 
maintaining correctness with two properties. 
Compared to state-of-the-art systems, 
\proj
reduces the execution time of GPM on large graphs and patterns 
from days to a few hours with low programming effort.

\bibliographystyle{ACM-Reference-Format}
\bibliography{ref}

\end{document}